\documentclass[aps,prd,twocolumn,nofootinbib,showpacs,amsmath,amssymb,floatfix,eqsecnum]{revtex4-1}

\usepackage{graphicx,color,framed}
\usepackage{times}
\usepackage{enumerate}
\usepackage{lipsum}
\usepackage{slashed}
\usepackage{color}
\usepackage{amsfonts}
\usepackage{amssymb}
\usepackage{mathrsfs}
\usepackage{hyperref}

\hypersetup{
    colorlinks=true, %set true if you want colored links
    linktoc=all,     %set to all if you want both sections and subsections linked
    linkcolor=blue,  %choose some color if you want links to stand out
}

\def \beq {\begin{equation}}
\def \eeq {\end{equation}}
\def \beqa {\begin{eqnarray}}
\def \eeqa {\end{eqnarray}}
\def \bseq {\begin{subequations}}
\def \eseq {\end{subequations}}
\newcommand \dg {\dagger}

\newcommand \al {\alpha}

\newcommand \ran {\rangle}
\newcommand \lan {\langle}

\newcommand \ep {\epsilon}

\newcommand \pd {\partial}

\newcommand \mb {\mathbf}

\newcommand \nnb {\nonumber}
\newcommand \ov {\overline}

\newcommand \vphi {\varphi}
\newcommand \D {\mathcal{D}}

\begin{document}

\title{Geometric quench in the fractional quantum Hall effect: exact solution in quantum Hall matrix models and comparison with bimetric theory}

\author{Matthew F. Lapa}
\email[email address: ]{mlapa@uchicago.edu}
\affiliation{Kadanoff Center for Theoretical Physics, University of Chicago, Illinois 60637, USA}

\author{Andrey Gromov}
\email[email address: ]{gromovand@uchicago.edu}
\affiliation{Kadanoff Center for Theoretical Physics, University of Chicago, Illinois 60637, USA}
\affiliation{Department of Physics, University of California, Berkeley, California 94720, USA}
\affiliation{Materials Sciences Division, Lawrence Berkeley National Laboratory, Berkeley, California 94720, USA}

\author{Taylor L. Hughes}
\email[email address: ]{hughest@illinois.edu}
\affiliation{Department of Physics and Institute for Condensed Matter Theory, University of Illinois at Urbana-Champaign, Urbana, IL, 61801-3080, USA}

%\date{\today}

\begin{abstract}

We investigate the recently introduced geometric quench protocol for fractional quantum Hall (FQH) states within the framework of 
exactly solvable quantum Hall matrix models. In the geometric quench protocol a FQH state is subjected 
to a sudden change in the ambient geometry, which introduces anisotropy into the system. We formulate this quench in the matrix models 
and then we solve exactly for the post-quench dynamics of the system and the quantum fidelity (Loschmidt 
echo) of the post-quench state. Next, we explain how to define a spin-2 collective variable $\hat{g}_{ab}(t)$ in the matrix models, and we 
show that for a weak quench (small anisotropy) the dynamics of $\hat{g}_{ab}(t)$ agrees with the dynamics of the intrinsic metric governed 
by the recently discussed bimetric theory of FQH states. We also find a modification of the bimetric theory such that the predictions of the
modified bimetric theory agree with those of the matrix model for arbitrarily strong quenches. 
Finally, we introduce a class of higher-spin collective variables for the matrix model, which are related to generators of the $W_{\infty}$ 
algebra, and we show that the geometric quench induces nontrivial dynamics for these variables.

\end{abstract}


\maketitle

\section{Introduction}

Topological phenomena in gapped fractional quantum Hall (FQH) states, such as anyonic excitations, robust edge modes, and
ground state degeneracy on closed manifolds, are well-described by Chern-Simons topological quantum field 
theories~\cite{wen-book}. 
These theories apply in the limit in which the bulk energy gap is sent to infinity and so, by their very nature, they are
incapable of describing the dynamics of gapped excitations in FQH states.
Nevertheless, FQH states support a bulk gapped collective excitation known as the \emph{magneto-roton}, or 
\emph{Girvin-MacDonald-Platzman} (GMP) mode~\cite{GMP}. For small wavevectors $\mb{k}$ the GMP mode is characterized 
by a definite angular momentum equal to $2\hbar$, i.e., the GMP mode is a ``spin-2'' mode near $\mb{k}=0$. 
Recently, a new effective ``bimetric'' field theory was developed~\cite{GGB,gromov-son} to describe the
gapped dynamics of this spin-2 mode.
The fundamental degree of freedom in this theory is a dynamical unimodular metric $\hat{g}_{ab}(\mb{x},t)$, 
and the gapped fluctuations of this metric, which have spin-2, correspond to the dynamics of the GMP mode near $\mb{k}=0$. 
The development of the 
bimetric theory relied on the extensive body of work on geometry~\cite{AG,cho2014,ferrari-klevtsov,BR1,BR2,CLW,framing,gromov2014density,KN,
haldane2009,haldane2011,park-haldane,YCF-nematic,haldane-anisotropic, gromov2015thermal, gromov2016boundary, can2014fractional, douglas2010bergman, klevtsov2015geometric, schine2016synthetic, schine2018measuring, maciejko2013field} and Hall 
viscosity~\cite{ASZ,levay,avron1998odd,TV1,read2009,TV2,haldane2009,haldane2011,read-rezayi,HLF2011,
hoyos-son,bradlyn2012,park-haldane} in quantum Hall states from the past two decades. 

Given the existence of interesting gapped excitations in FQH states, it is natural to try to engineer a situation in which the 
gapped dynamics of FQH states could be observed, either in numerical simulations or in experiments. 
With this goal in mind, a quantum quench protocol for FQH states, 
dubbed a ``geometric quench'', was introduced in Ref.~\onlinecite{LGP}. 
This geometric quench is designed for the express purpose of exciting the (neutral) 
gapped excitations in FQH systems, and can be summarized briefly as follows. First, we prepare the system
in an isotropic FQH ground state $|\psi_0\ran$ of an isotropic Hamiltonian $H_0$. Next, we suddenly change the Hamiltonian to 
incorporate some anisotropy, $H_0\to H'$. Finally, we evolve the initial state forward in time using the new anisotropic Hamiltonian,
$|\psi(t)\ran= e^{-i\frac{H' t}{\hbar}}|\psi_0\ran$. 

The authors of Ref.~\onlinecite{LGP} investigated this geometric quench in two ways. First, they studied the quench analytically
using the aforementioned bimetric theory. Second, they studied the quench numerically using the recently introduced anisotropic Haldane 
pseudopotentials~\cite{Papic2017}. For quadropolar anisotropy parametrized by 
a constant unimodular metric $g_{ab}$, this quench was shown to excite the gapped spin-2 mode near $\mb{k}=0$ 
(i.e., the small $\mb{k}$ limit of the GMP mode). 
In addition, the dynamics of this mode in the case of \emph{small} anisotropy was shown to be well-described by bimetric theory. 
Ref.~\onlinecite{LGP} also considered quenches with more complicated
anisotropy, and these quenches were shown to excite exotic higher-spin modes, which have a larger excitation gap 
than the spin-2 mode. The existence of such higher-spin excitations in the FQH effect has been anticipated since early
work on infinite-dimensional $W_{\infty}$ symmetry in FQH states~\cite{CTZ,karabali1994,karabali1994-2,flohr1994,CTZ-2}. 

Our goal in this paper is to study the geometric quench protocol in more detail. To do so we consider this quench in the
context of exactly solvable \emph{matrix models} of FQH states. The exact solubility
of these matrix models allows us to make significant analytical progress in studying the geometric quench. 
We focus most of our discussion on the matrix model for Laughlin states,
known as the Chern-Simons matrix model (CSMM). The CSMM was introduced by Polychronakos~\cite{P1}, who proposed it as a 
concrete regularization of Susskind's noncommutative Chern-Simons theory of the Laughlin states~\cite{susskind}, and
the CSMM and noncommutative Chern-Simons theory were subsequently studied by many 
authors~\cite{P3,MP,HVR,karabali-sakita1,karabali-sakita2,hansson2001,fradkin-NCCS,hansson2003,cappelli2005,tong-turner2015}. 
We will also explain how our results for the Laughlin states extend to a matrix model for the Blok-Wen series~\cite{blok-wen} 
of non-Abelian FQH states. 
This non-Abelian matrix model was introduced and studied in detail in Refs.~\onlinecite{tong2016,dorey2016matrix} (this model
also appeared in Ref.~\onlinecite{MP}, but was given a different physical interpretation in that reference). 
In Refs.~\onlinecite{lapa2018,LHTT}, it was shown that the matrix models accurately capture the geometric properties of the FQH 
states they describe. In particular, the correct value of the guiding center Hall viscosity of these FQH states can 
be recovered from the matrix model descriptions (see Refs.~\onlinecite{haldane2009,haldane2011,park-haldane}
for the concept of Landau orbit vs. guiding center contributions to the Hall viscosity). The fact that the
CSMM and its non-Abelian generalizations accurately describe the geometric response of FQH states suggest that these
models are ideal testing grounds for the geometric quench of Ref.~\onlinecite{LGP}.

In this paper we formulate the geometric quench protocol in the CSMM for the case of quadropolar anisotropy parametrized
by a constant unimodular metric $g_{ab}$. We then solve exactly for the post-quench state $|\psi(t)\ran$ and
compute the \emph{quantum fidelity} $|\lan\psi_0|\psi(t)\ran|^2$ (also known as the \emph{Loschmidt echo}). We also
define and compute the exact dynamics of a spin-2 collective variable that naturally emerges in the CSMM. We denote
this collective variable by $\hat{g}_{ab}(t)$ because, as we show in the paper, this
quantity is the analogue in the CSMM of the dynamical metric in bimetric theory. We show that
$\hat{g}_{ab}(t)$ undergoes nonlinear oscillations after the quench, with a period set by the gap $E_2$ for spin-2
excitations in the CSMM. In the case of \emph{small} anisotropy, we show that the dynamics of $\hat{g}_{ab}(t)$ in the
CSMM coincides with the post-quench dynamics predicted by bimetric theory in Ref.~\onlinecite{LGP}. 
We also generalize these results to the non-Abelian matrix model of Refs.~\onlinecite{tong2016,dorey2016matrix}. 
These results imply that the quantum Hall matrix models can describe the numerical data of Ref.~\onlinecite{LGP} for small 
anisotropy just as well as bimetric theory. 

We then explore the connection between the matrix models and bimetric theory in more detail, and we show that there
exists a modified potential energy term for bimetric theory such that the predictions of the matrix models for the geometric
quench \emph{exactly} match the predictions of bimetric theory with the alternative potential energy term. 
Finally, in the last part of the paper we define a set of higher-spin collective variables for the CSMM and discuss their relation
to previous work on higher-spin operators and $W_{\infty}$ symmetry in the CSMM. We then show that the geometric
quench considered in this paper induces nontrivial dynamics for these higher-spin variables. 

The CSMM is closely related to the Calogero model of interacting particles in one dimension (see Ref.~\onlinecite{P1} for
the connection). Consequently, there is a relation between the geometric quench in the CSMM and the quench of the
harmonic trap frequency in the Calogero model that was considered in Ref.~\onlinecite{rajabpour2014}. The main difference
between the geometric quench for the CSMM and the work of Ref.~\onlinecite{rajabpour2014} is that, in the language of the
Calogero model, the geometric quench of the CSMM corresponds to a \emph{simultaneous} quench of both the harmonic trap 
frequency \emph{and} the mass of the Calogero particles (note that in the Calogero Hamiltonian the mass parameter
appears as a coefficient in the kinetic energy term \emph{and} the interaction term). 
Thus, the dynamics induced by the geometric quench in the CSMM is 
qualitatively distinct from that studied in Ref.~\onlinecite{rajabpour2014}. Another similar quench protocol was discussed in 
\cite{franchini2015universal}, where the harmonic trap frequency was quenched simultaneously with the interaction strength.

This paper is organized as follows. In Sec.~\ref{sec:CSMM} we review the CSMM and introduce various important variables
and notation. In Sec.~\ref{sec:quench-solution} we formulate and solve the geometric quench in the CSMM, and
extend those results to the non-Abelian matrix model. In
Sec.~\ref{sec:bimetric} we give a detailed comparison of the predictions of the CSMM and bimetric theory, and we also discuss
the new potential energy term for bimetric theory that we mentioned in the previous paragraph. 
In Sec.~\ref{sec:higher-spin} we introduce a set of higher-spin collective variables for the CSMM, and we calculate their
post-quench dynamics. Sec.~\ref{sec:conclusion} presents our conclusions. Finally, several important formulas are contained
in Appendices~\ref{app:SU11}, \ref{app:big-formulas}, and \ref{app:W}.

\section{Review of the Chern-Simons matrix model (CSMM)}
\label{sec:CSMM}

\subsection{Physical meaning of the model and summary of notation}

In this section we give a lightning review of the CSMM and its quantization. We also highlight some specific properties of the
quantum ground state of the CSMM which we use later in the paper in the solution of the geometric quench. For more details
on this model and its physical interpretation we refer the reader to the original work \cite{P1}, and to \cite{lapa2018,LHTT} for a 
recent discussion in the context of geometric response of quantum Hall states (our notation is essentially the same as
Ref.~\onlinecite{LHTT}). 

The degrees of freedom in the CSMM consist of two $N\times N$ Hermitian matrices $X^a(t)$, $a=1,2$, a complex length $N$ 
vector $\vphi(t)$, and an additional $N\times N$ Hermitian matrix $A_0(t)$ which is a $U(N)$ gauge field. All of these degrees 
of freedom are functions of time $t$. We denote the matrix elements of the
matrix degrees of freedom by ${(X^a)^j}_k$, $j,k=1,\dots,N$ (and likewise for $A_0$), and the components of $\vphi$
by $\vphi^j$, $j=1,\dots,N$. 

The physical meaning of the CSMM can be briefly summarized as follows. The starting point for this interpretation is Susskind's
noncommutative Chern-Simons theory description of the Laughlin states~\cite{susskind}. In that description a quantum Hall 
state is modeled as a fluid on the ``noncommutative plane", a deformation of the two-dimensional plane $\mathbb{R}^2$ in 
which the coordinates $x^a$ are promoted to operators $\hat{x}^a$ which obey a nontrivial commutation relation 
$[\hat{x}^1,\hat{x}^2]=i\theta$, where $\theta$ is a constant with units of length squared. In Susskind's model $\theta$ is 
quantized as
\beq
	\theta= \ell_B^2 m\ , m\in \mathbb{Z}\ ,
\eeq
where  $\ell_B^2=\frac{\hbar}{eB}$ is the square of the magnetic length\footnote{We use a convention in which 
electrons have charge $-e <0$, and we choose a constant magnetic field with strength $B>0$ (i.e., pointing in the positive
$z$ direction).}. The integer $m$, which we take to be positive, is related to the filling fraction of the Laughlin state by
\beq
	\nu=\frac{1}{m}\ .
\eeq
This can be seen from the fact that the density of the fluid in Susskind's model is related to $\theta$ by
\beq
	\ov{\rho}= \frac{1}{2\pi\theta}=\frac{1}{2\pi \ell_B^2 m}\ ,
\eeq
which is exactly the mean density of the $\nu=\frac{1}{m}$ Laughlin state. The physical interpretation of the parameter
$\theta$ is that $2\pi\theta$ is the area occupied by a single electron in Susskind's model. In \cite{susskind} it was 
argued that due to the finite value of the parameter $\theta$, the noncommutative Chern-Simons theory accurately 
captures the ``granularity" of a fluid composed of discrete particles (which are electrons in this case).

The CSMM can be viewed as a regularization of Susskind's noncommutative Chern-Simons theory. While the latter theory 
describes a constant density fluid occupying the entire noncommutative plane, the CSMM describes a finite droplet 
of fluid on the noncommutative plane consisting of $N$ electrons. Indeed, the eigenvalues of the matrices 
$X^a$ in the CSMM can be interpreted as the coordinates of electrons on the plane. Since the matrices $X^a$ do not
commute with each other in the CSMM (i.e., they are not simultaneously diagonalizable), the electrons described by the 
CSMM still live on the noncommutative plane. For further details on the physical interpretation of the CSMM we refer the
reader to \cite{susskind,P1,lapa2018}.

Before moving on, we summarize our notations. 
When the matrix model is quantized, the matrix elements of $X^a$ and $A_0$, as well as the components of $\vphi$, become 
operators on a quantum Hilbert space. In what follows we reserve the symbol ``$\dg$" to
denote Hermitian conjugation of quantum operators. For classical matrix and vector degrees of freedom we use a superscript 
``$T$" to denote a transpose and an overline to denote complex conjugation.
We also use the notation $[\cdot,\cdot]_M$ to 
denote the commutator of classical matrix degrees of freedom (``$M$" stands for matrix). The notation $[\cdot,\cdot]$ without
any subscript will be used for the commutator of quantum operators. Finally, $\text{Tr}\{\cdot\}$ always 
denotes the trace of classical matrices.

\subsection{CSMM and its quantization}

The action for the CSMM has the form\footnote{We use a summation convention in which we sum over any index which
appears once as a subscript and once as a superscript in any expression.}
\begin{align}
	S_0 &= -\frac{eB}{2}\int_0^T dt\ \text{Tr}\Big\{ \ep_{ab}X^a \D_0 X^b + 2\theta A_0 \nnb \\
+&\ \omega \delta_{ab}X^a X^b \Big\} + i \int_0^T dt\ \ov{\vphi}^{T}\D_0 \vphi\ ,
\end{align}
where the covariant derivatives $\mathcal{D}_0 X^b$ and $\mathcal{D}_0\vphi$ are defined as
\begin{subequations}
\beqa
	\D_0 X^b &=& \dot{X}^b- i[A_0,X^b]_M \\
	\D_0\vphi &=& \dot{\vphi}- i A_0 \vphi\ ,
\eeqa
\end{subequations}
and the dot denotes an ordinary time derivative. Here we work on a time interval $t\in[0,T)$, and we impose periodic boundary
conditions in time on all degrees of freedom. This turns the time-direction into a circle of circumference $T$, which we denote by
$S^1_T$. Just as in Susskind's model, the parameter $\theta$ is quantized as $\theta= \ell_B^2 m\ ,\ m\in\mathbb{Z}$ and
we again choose $m>0$. In this case the CSMM describes the Laughlin state with $\nu=\frac{1}{m}$.

The quantization rule for $\theta$ comes from requiring the exponential $e^{i\frac{S_0}{\hbar}}$ of the 
action to be invariant under large $U(N)$ gauge transformations. The action $S_0$ is nearly invariant under the
$U(N)$ gauge transformation
\begin{subequations}
\beqa
	X^a &\to& VX^a \ov{V}^T \\
	A_0 &\to& VA_0 \ov{V}^T + i V \dot{\ov{V}}^T \\
	\vphi &\to& V\vphi\ ,
\eeqa
\end{subequations}
where $V(t)$ is a time-dependent $U(N)$ matrix. However, the term in the Lagrangian proportional to $\text{Tr}\{A_0\}$
spoils this invariance. This is because of the existence of large gauge transformations in which the map $V: S^1_T \to U(N)$
corresponds to a nontrivial element of the group $\pi_1(U(N))=\mathbb{Z}$. Requiring invariance of 
$e^{i\frac{S_0}{\hbar}}$ under these large gauge transformations then gives the quantization rule for $\theta$.

In the CSMM the gauge field $A_0$ enforces the constraint ($\mathbb{I}$ is the $N\times N$ identity matrix)
\beq
	G := 	ieB[X^1,X^2]_M+eB\theta\mathbb{I} - \vphi\ov{\vphi}^T= 0 \ , \label{eq:constraint}
\eeq
and in the $A_0=0$ gauge the Hamiltonian takes the form
\beq
	H_0= \frac{eB\omega}{2}\text{Tr}\{\delta_{ab}X^a X^b\}\ .
\eeq
This Hamiltonian represents a harmonic trap for the noncommutative fluid described by the CSMM, and the strength of this
trap is set by the frequency $\omega$.

To quantize the model it is convenient to define a set of real scalar variables which serve to completely specify the matrices
$X^a$. In the quantized model these variables then become Hermitian operators. To define these real scalar variables we 
introduce a basis $T^A$, $A=0,\dots,N^2-1,$ of generators of the Lie algebra of $U(N)$ in the fundamental representation.
Thus, $T^A$ are $N\times N$ Hermitian matrices, and we assume they are normalized so that
$\text{Tr}\{T^A T^B\}= \delta^{AB}$. A concrete choice for 
the generators $T^A$ is to choose $T^0= \frac{\mathbb{I}}{\sqrt{N}}$, and so $T^0$ is the generator
of the $U(1)$ part of $U(N)$. For $A\neq 0$ we choose
$T^A= \sqrt{2} t^A$, where $t^A$ are a basis of conventionally normalized generators of $SU(N)$ which satisfy 
$\text{Tr}\{t^A t^B\}=\frac{\delta^{AB}}{2}$ and $[t^A,t^B]_{M}= i \sum_C f^{ABC} t^C$, where $f^{ABC}$ 
are the structure constants of $SU(N)$ (we do not need to know the exact form of $f^{ABC}$ in this paper).
Using this basis we then parametrize $X^a(t)$ as
\beq
	X^a(t)= \sum_{A=0}^{N^2-1} x^a_A(t) T^A\ ,
\eeq
where we have introduced $2N^2$ real scalar variables $x^a_A(t)$.
 In the quantized CSMM these scalar variables obey the commutation relations
\beq
	[x^a_A, x^b_B]= i\ell_B^2 \ep^{ab}\delta_{AB}\ ,
\eeq
where are very similar to the commutation relations of \emph{guiding center} coordinates in the quantum Hall problem.

Using these new scalar variables we define the oscillator variables
\beq
	z_A = \frac{1}{\ell_B\sqrt{2}}( x^1_A + i x^2_A)\ ,
\eeq
and $z^{\dg}_A = \frac{1}{\ell_B\sqrt{2}}( x^1_A - i x^2_A)$. We also define 
\beq
	b^j= \frac{1}{\sqrt{\hbar}}\vphi^j \ ,
\eeq
and $b^{\dg}_j= \frac{1}{\sqrt{\hbar}}\ov{\vphi}_j$ (here $\ov{\vphi}_j$ are the components of the row vector 
$\ov{\vphi}^T$). In the quantized CSMM these variables all obey the
harmonic oscillator commutation relations
\begin{subequations}
\beqa
	\left[z_A, z^{\dg}_B\right] &=& \delta_{AB} \\
	\left[ b^j, b^{\dg}_k\right]&=& \delta^j_k\ .
\eeqa
\end{subequations}
For later use we also define the matrix-valued operators $Z^{\pm}$ whose matrix elements are given by
\begin{subequations}
\beqa
	{(Z^{-})^j}_k &=& \sum_{A=0}^{N^2-1} z_A {(T^A)^j}_k \\
	{(Z^{+})^j}_k &=& \sum_{A=0}^{N^2-1} z^{\dg}_A {(T^A)^j}_k\ .
\eeqa
\end{subequations}
The commutation relations of $z_A$ and $z^{\dg}_B$ then imply that 
\beq
	\left[{(Z^{-})^j}_k, {(Z^{+})^{\ell}}_m  \right] = \delta^j_m \delta^{\ell}_k\ . 
\eeq

If we quantize the CSMM in the $A_0=0$ gauge, then gauge invariance requires that all states in the physical Hilbert
space of the model be annihilated by the matrix elements ${G^j}_k$ of the constraint $G$ from Eq~\eqref{eq:constraint}.
A useful way to think about these constraints is to define a new set of constraints by taking the trace with the 
$U(N)$ generators $T^A$, i.e., we define new constraints $G^A := \text{Tr}\{G T^A\}$. Let $|\text{phys}\ran$ denote a state in 
the physical Hilbert space of the model. Then the constraints $G^A|\text{phys}\ran=0$ for $A\neq 0$ imply that all physical states
transform as singlets under the $SU(N)$ part of $U(N)$. The remaining constraint $G^0|\text{phys}\ran=0$ can be shown 
to reduce to 
\beq
	b^{\dg}_j b^j |\text{phys}\ran= N(m-1)|\text{phys}\ran\ .
\eeq
This constraint implies that all physical states carry a total charge of $N(m-1)$ under the $U(1)$ part of $U(N)$. Note also
that $G^A|\text{phys}\ran=0$ for all $A$ implies that ${G^j}_k|\text{phys}\ran=0$ for all $j,k$, since the ${G^j}_k$ are linear 
combinations of the $G^A$.

Let $|0\ran$ be the Fock vacuum state which is annihilated by the $z_A$ and $b^j$ operators. 
Then a complete basis of physical states for the CSMM consists of the states~\cite{HVR}
\begin{align}
	|\{c_1,\dots,c_N\}\ran &= \nnb \\
	 \text{Tr}\{Z^{+}\}^{c_1}&\text{Tr}\{(Z^{+})^2\}^{c_2}\cdots\text{Tr}\{(Z^{+})^N\}^{c_N}|\psi_0\ran  \label{eq:excited-states}
\end{align}
where $c_j \in \mathbb{N}$ for $j=1,\dots,N$, and 
\beq
	|\psi_0\ran= \left(\ep^{j_1\cdots j_N}b^{\dg}_{j_1}[b^{\dg}Z^{+}]_{j_2}\cdots[b^{\dg}(Z^{+})^{N-1}]_{j_N}\right)^{(m-1)}|0\ran\ .\label{eq:CSMM-ground-state}
\eeq 
In the $A_0=0$ gauge the CSMM Hamiltonian can be rewritten in the form
\beq
	H_0= \hbar\omega\frac{N^2}{2} + \hbar\omega \sum_{A=0}^{N^2-1}z^{\dg}_A z_A\ ,
\eeq
which is equal to a constant plus a term proportional to the total number operator for the $z_A$ oscillators. 
From this it is clear that the lowest energy physical state is $|\psi_0\ran$, with energy
\beq
	E_0 = \hbar\omega\left[\frac{1}{2}mN^2 +\left(\frac{1-m}{2}\right)N\right]\ .
\eeq
The other states $|\{c_1,\dots,c_N\}\ran$ can be seen to have an energy of
\beq
	E(\{c_1,\dots,c_N\})= E_0 + \hbar\omega \sum_{j=1}^N c_j j\ .
\eeq
For later use we also define a dimensionless ground state energy $\ep_0$ by
\beq
	\ep_0 := \frac{E_0}{\hbar \omega}\ .
\eeq

We also mention here that the angular momentum operator $L_z$ for the CSMM takes the form
\beq
	L_z= -\frac{eB}{2}\text{Tr}\{\delta_{ab}X^a X^b\}\ . \label{eq:Lz}
\eeq
In particular, it is clear that $L_z= -\frac{1}{\omega}H_0$. It follows that the state $|\{c_1,\dots,c_N\}\ran$ has angular
momentum
\beq
	L_z(\{c_1,\dots,c_N\})= -\hbar\ep_0 - \hbar\sum_{j=1}^N c_j j\ .
\eeq

\subsection{$sl(2,\mathbb{R})$ generators}

The matrices $X^a$ can be interpreted as Lagrangian coordinates for a fluid on the noncommutative plane \cite{susskind}. 
To investigate the response of this fluid to changes in the geometry, we need to identify the operators which 
generate area-preserving diffeomorphisms (APDs) of the fluid coordinates. The group $\text{SDiff}(\mathbb{R}^2)$ of
APDs of the plane is an infinite-dimensional group whose elements are (smooth, invertible) 
functions $\eta: \mathbb{R}^2\to\mathbb{R}^2$ which
preserve the volume form $\text{vol}=dx^1\wedge dx^2$ on $\mathbb{R}^2$, i.e., $\eta^*\text{vol}=\text{vol}$,
where $\eta^*$ denotes the pullback along the map $\eta$. This group has a finite-dimensional subgroup isomorphic
to $SL(2,\mathbb{R})$ which consists of the functions $\eta$ of the form
\beq
	\eta^a(x)= {A^a}_b x^b\ ,
\eeq
where ${A^a}_b$ are the components of a $2\times 2$ real matrix $A$ with determinant $1$, i.e., an element of 
$SL(2,\mathbb{R})$. Note here that ${A^a}_b$ has no $x$ dependence. One can think of this subgroup of 
$\text{SDiff}(\mathbb{R}^2)$ as being equal to the subset of APDs which are uniform in space.

%\AG{I don't think that APD is the right name for $\mathsf{\Lambda}^{ab}$ since the former is an infinite-dimenstional algebra, while the latter operators form $sl(2,R)$. The idea is that all of the higher spin fields combined form APD} \textbf{MFL: OK, that's an important distinction. Let's talk about what to call them then. We did call
%them APD generators in our previous papers, and I think we did that because that is what Haldane called them in his papers}. 
It was shown in \cite{lapa2018} that the operators which generate these $SL(2,\mathbb{R})$ transformations for the
matrix coordinates $X^a$ are\footnote{Note that in \cite{lapa2018,LHTT} 
these operators were referred to as ``area-preserving
deformation" generators. Here we refer to them as $sl(2,\mathbb{R})$ generators to make the connection with
the full group of area-preserving diffeomorphisms of $\mathbb{R}^2$ more precise.}
\beq
	\mathsf{\Lambda}^{ab}=  \frac{1}{4\ell_B^2}\sum_{A=0}^{N^2-1} \{ x^a_A, x^b_A\}\ ,
\eeq
where $\{\cdot,\cdot\}$ denotes an anti-commutator, and one can check that these operators obey,
\beq
	[\mathsf{\Lambda}^{ab},\mathsf{\Lambda}^{cd}] = \frac{i}{2}\left(\ep^{bc}\mathsf{\Lambda}^{ad} + \ep^{bd}\mathsf{\Lambda}^{ac} + \ep^{ac}\mathsf{\Lambda}^{bd}+\ep^{ad}\mathsf{\Lambda}^{bc}  \right)\ , \label{eq:SL2R}
\eeq
which are the commutation relations for the Lie algebra $sl(2,\mathbb{R})$.

Finite $SL(2,\mathbb{R})$ transformations of the noncommutative coordinates are implemented by conjugation by a unitary 
operator $U(\al)= e^{i\al_{ab}\mathsf{\Lambda}^{ab}}$, where $\al_{ab}$ is 
a constant symmetric matrix which parametrizes the deformation. To first order in $\al_{ab}$ we have (for all $j,k$)
\beq
	U(\al) {(X^a)^j}_k U(\al)^{\dg}= {(X^a)^j}_k + \ep^{ab}\al_{bc}{(X^c)^j}_k+\cdots
\eeq
Since the operators $U(\al)$ act identically on all matrix elements ${(X^a)^j}_k$ of the noncommutative coordinates 
$X^a$, the operators $\mathsf{\Lambda}^{ab}$ can indeed be interpreted as generating $SL(2,\mathbb{R})$ 
transformations of the noncommutative coordinates $X^a$. 

It is convenient to introduce another basis for the generators of $sl(2,\mathbb{R})$, which have the form
%the set of shear generators
\label{eq:su11-APD}
\beqa
K_0 &=& \frac{1}{2}\Big(\mathsf{\Lambda}^{11} + \mathsf{\Lambda}^{22}\Big) \\
K_{-} &=& \frac{1}{2}\Big(\mathsf{\Lambda}^{11} - \mathsf{\Lambda}^{22}\Big) + i \mathsf{\Lambda}^{12} \\
K_{+}&=& (K_{-})^{\dg} .
\eeqa
This basis of generators obeys the algebra
\begin{subequations}
\beqa
	[K_0,K_{\pm}] &=& \pm K_{\pm} \\
	\left[K_{-},K_{+}\right] &=& 2K_0\ ,
\eeqa
\end{subequations}
and in this form the $sl(2,\mathbb R)$ algebra is also known as $su(1,1)$.

One fact which will be useful later in the paper is that $K_{-}$ annihilates the ground state $|\psi_0\ran$ of the 
original CSMM,
\beq
	K_{-}|\psi_0\ran= 0\ ,
\eeq
and this can be shown using a proof by contradiction. Suppose instead that $K_{-}|\psi_0\ran \neq 0$. Then, since 
$K_{-}$ is invariant under the $U(N)$ action in the CSMM (this can be seen by writing it as a trace, 
$K_{-}=\frac{1}{2}\text{Tr}\{ (Z^{-})^2\}$), the state $K_{-}|\psi_0\ran$ is also a
valid state in the physical Hilbert space of the matrix model. In addition, this
state has energy $E_0-2\hbar \omega$ for the Hamiltonian $H_0$. Therefore $K_{-}|\psi_0\ran$, if different from
zero, would be a new physical state of the CSMM with lower energy than the ground state $|\psi_0\ran$. This is a
contradiction since it is already known that $|\psi_0\ran$ has the lowest eigenvalue of $H_0$ among all of the
physical states of the model. Therefore it must be that $K_{-}|\psi_0\ran = 0$. Note that this proof also generalizes to a proof 
that $|\psi_0\ran$ is annihilated by all the $U(N)$-invariant operators $\text{Tr}\{ (Z^{-})^p\}$ for $p=1,\dots,N$, with 
$K_{-}$ corresponding to the case of $p=2$.

\subsection{Introducing anisotropy into the CSMM}

We now explain how to introduce anisotropy into the CSMM. One way to do this, following \cite{lapa2018}, 
is to deform the harmonic trap by 
replacing the Kronecker delta $\delta_{ab}$ with a constant unimodular metric $g_{ab}$ (i.e., a constant metric with
determinant equal to one). The nontrivial metric $g_{ab}$ represents some externally imposed anisotropy in the problem.
The action for this modified CSMM takes the form
\begin{align}
	S_g &= -\frac{eB}{2}\int_0^T dt\ \text{Tr}\Big\{ \ep_{ab}X^a \D_0 X^b + 2\theta A_0 \nnb \\
+&\ \omega g_{ab}X^a X^b \Big\} + i \int_0^T dt\  \ov{\vphi}^{T}\D_0 \vphi\ , \label{eq:CSMM-g}
\end{align}
in which the only change to the action is the replacement $\delta_{ab} \to g_{ab}$ in the harmonic potential term. In the 
$A_0=0$ gauge the Hamiltonian for this modified matrix model is
\beq
	H_g= \frac{eB\omega}{2}\text{Tr}\{g_{ab}X^a X^b\}\ .
\eeq

This model was solved in \cite{lapa2018} and we mention here some of the important properties of this model. First, 
the entire energy spectrum of this model is identical to that of the CSMM with $g_{ab}=\delta_{ab}$ (this statement is only
true because $g_{ab}$ has determinant one). In particular, the quantum ground state $|\psi_g\ran$ of this model has the same 
energy $E_0$ as the ground state $|\psi_0\ran$ of the original CSMM. In addition, the expectation value of the
$sl(2,\mathbb{R})$ generators $\mathsf{\Lambda}^{ab}$ in the state $|\psi_g\ran$ is given by
\beq
	\lan \psi_g |\mathsf{\Lambda}^{ab}|\psi_g\ran = \frac{\ep_0}{2}g^{ab}\ , \label{eq:metric-time-independent}
\eeq
where $g^{ab}$ is the inverse metric for $g_{ab}$. It was shown in \cite{lapa2018} that, as a consequence of this
relation, the Hall viscosity of this modified CSMM with $\theta=\ell_B^2 m$ is equal to the guiding center Hall viscosity of the
Laughlin $\nu=\frac{1}{m}$ state with guiding center metric $g_{ab}$~\cite{haldane2009,haldane2011}.

This calculation also suggests a way to define an intrinsic metric $\hat{g}_{ab}$ associated with any state
$|\psi\ran$ in the matrix model. We define $\hat{g}_{ab}$ by first defining its inverse
$\hat{g}^{ab}$ to be proportional to the expectation value $\lan\psi| \mathsf{\Lambda}^{ab}|\psi\ran$. In the special case
that $|\psi\ran$ is chosen to be the ground state $|\psi_g\ran$ of the CSMM with metric $g_{ab}$, Eq.~\eqref{eq:metric-time-independent} 
shows that the intrinsic metric $\hat{g}_{ab}$ associated with this state is locked to the externally imposed metric $g_{ab}$. 
Later in the paper we use the intuition 
provided by this example to define a time-dependent intrinsic
metric $\hat{g}_{ab}(t)$ in a time-dependent state $|\psi(t)\ran$ obtained after performing a geometric quench in the CSMM.

Finally we note that the Hamiltonian $H_g$ can be written in terms of the
$sl(2,\mathbb{R})$ generators as
\beq
	H_g= \hbar\omega g_{ab}\mathsf{\Lambda}^{ab}\ . \label{eq:Ham-g-APD}
\eeq
In this form the Hamiltonian of the CSMM resembles the Hamiltonian of the bimetric theory, as we discuss later.

% \subsection{$su(1,1)$ algebra}

% In this subsection we define a set of operators which will prove very useful in the rest of the paper. 
% Using the set of oscillator variables $z_A, z^{\dg}_A$, $A=0,\dots,N^2-1$, which obey the standard harmonic
% oscillator commutation relations $[z_A,z^{\dg}_B]= \delta_{AB}$, we can define a set of operators
% \begin{subequations}
% \beqa
% 	K^{+-} &=& \frac{1}{2}\sum_{A=0}^{N^2-1}\left( z^{\dg}_A z_A + \frac{1}{2}\right) \\
% 	K^{++} &=& \frac{1}{2}\sum_{A=0}^{N^2-1} (z^{\dg}_A)^2 \\
% 	K^{--} &=& \frac{1}{2}\sum_{A=0}^{N^2-1} (z_A)^2 \ ,
% \eeqa
% \end{subequations}
% and these operators obey the Lie algebra
% \begin{subequations}
% \beqa
% 	[K^{+-},K_{\pm}] &=& \pm K_{\pm} \\
% 	\left[K^{--},K^{++}\right] &=& 2K^{+-}\ .
% \eeqa
% \end{subequations}
% This Lie algebra is usually called $su(1,1)$, the Lie algebra of the group $SU(1,1)$, but $su(1,1)$ is in fact isomorphic to 
% $sl(2,\mathbb{R})$. This allows us to express the 
% generators $\mathsf{\Lambda}^{ab}$ in terms of the $su(1,1)$ generators as
% \begin{subequations}
% \label{eq:su11-APD}
% \beqa
% 	\mathsf{\Lambda}^{11} &=& K^{+-} + \frac{1}{2}K^{++} + \frac{1}{2}K^{--} \\
% 	\mathsf{\Lambda}^{22} &=& K^{+-} - \frac{1}{2}K^{++} - \frac{1}{2}K^{--}  \\
% 	\mathsf{\Lambda}^{12} &=& \mathsf{\Lambda}^{21}= \frac{-i}{2}(K^{--}-K^{++})\ .
% \eeqa
% \end{subequations}

\section{Geometric quench in the CSMM and its exact solution}
\label{sec:quench-solution}

In this section we formulate the geometric quench protocol in the CSMM, and we also define a time-dependent intrinsic
metric in the post-quench state. We then present the exact solution for the post-quench state and the time-dependent
intrinsic metric. We show that our results agree with the results obtained in \cite{LGP} using bimetric theory in the limit of 
small anisotropy (we give a more detailed comparison with the bimetric theory
results later in Sec.~\ref{sec:bimetric}). Finally, at the end of this section we show how our results for the CSMM can be extended 
to the case of the non-Abelian matrix model of Refs.~\onlinecite{tong2016,dorey2016matrix}. 

\subsection{Geometric quench protocol in the CSMM}

The geometric quench of a FQH state was introduced in Ref.~\onlinecite{LGP} and consists of a sudden change in the
background geometry in a FQH system. This quench can be formulated in the CSMM as follows. We start with the system in the 
ground state $|\psi_0\ran$ of the original CSMM. Then, at time $t=0$, we suddenly introduce anisotropy into the
system by replacing $\delta_{ab} \to g_{ab}$ in the harmonic trap term of the CSMM (we still take $g_{ab}$ to be a 
constant unimodular metric). As result, the initial state 
$|\psi_0\ran$ evolves in time under the influence of the Hamiltonian $H_g$ of the CSMM with nontrivial metric
$g_{ab}$ from Eq.~\eqref{eq:CSMM-g}. Mathematically, the post-quench state at time $t$ is related to the initial 
state as
\beq
	|\psi(t)\ran= e^{-i \frac{H_g t}{\hbar}}|\psi_0\ran\ .
\eeq
One of our main results in this section is an explicit expression for the post-quench state $|\psi(t)\ran$.

Given the post-quench state $|\psi(t)\ran$, we can define a time-dependent intrinsic metric using 
Eq.~\eqref{eq:metric-time-independent} as a guide. We denote this metric by $\hat{g}_{ab}(t)$ and we define it by first
defining its inverse $\hat{g}^{ab}(t)$ as
\beq
	\hat{g}^{ab}(t):= \frac{2}{\ep_0}\lan \psi(t)|\mathsf{\Lambda}^{ab}|\psi(t)\ran\ . \label{eq:metric-time-dependent}
\eeq
The normalization factor here can be understood by comparison with Eq.~\eqref{eq:metric-time-independent}, and with
this normalization $\hat{g}_{ab}(t)$ will also be a unimodular metric (this will be verified by an explicit computation). 
The ``dynamical metric'' $\hat{g}_{ab}(t)$ is a spin-$2$ collective variable, which (partially) characterizes the many-body 
dynamics of the CSMM. 

For easy comparison with Ref.~\onlinecite{LGP}, we choose the anisotropy metric $g_{ab}$ to be of the form
\beq
	g= \begin{pmatrix}
	e^A & 0 \\
	0 & e^{-A}
\end{pmatrix}\ , \label{eq:anisotropy-g}
\eeq
where $A$ is a real parameter which determines the anisotropy ($g_{ab}$ are the components of the matrix $g$). 
This choice of metric stretches the system along one axis
(the $x^1$-axis for $A>0$), while squashing the system along the other axis. The fact that $g_{ab}$ is diagonal means that
there is no additional rotation off of the main coordinate axes. This choice of $g_{ab}$ makes our calculations in this
section slightly easier, however, the case of a non-diagonal $g_{ab}$ can be dealt with using the same methods. 

To make contact with Ref.~\onlinecite{LGP} we also parametrize the dynamical metric $\hat{g}_{ab}(t)$ using a
a real parameter $Q(t)\geq 0$ and a real phase $\phi(t)$. In this parametrization the metric takes the form
considered in \cite{LGP},
\beq
	\hat{g}=\begin{pmatrix}
	\cosh(Q)+\cos(\phi)\sinh(Q) & \sin(\phi)\sinh(Q) \\
	\sin(\phi)\sinh(Q)  & \cosh(Q)-\cos(\phi)\sinh(Q)
\end{pmatrix}\ . \label{eq:LGP-metric}
\eeq
One can check that this does indeed define a unimodular metric.
We now proceed with the exact
calculations of $|\psi(t)\ran$ and $\hat{g}_{ab}(t)$.

%We also choose to parametrize the dynamical metric $\hat{g}_{ab}(t)$ using
%a time-dependent complex parameter $\beta(t)$ satisfying $|\beta(t)|<1$. In terms of $\beta(t)$, $\hat{g}_{ab}(t)$
%takes the form
%\beq
%	\hat{g}= \frac{1}{1-|\beta|^2}\begin{pmatrix}
%	(1+\beta)(1+\ov{\beta}) & -i (\beta-\ov{\beta}) \\
%	-i (\beta-\ov{\beta})  & (1-\beta)(1-\ov{\beta})
%\end{pmatrix}\ . \label{eq:unimodular-metric}
%\eeq

\subsection{Post-quench state and the quantum fidelity (Loschmidt echo)}

To determine the post-quench state $|\psi(t)\ran$ we first note that using
expression \eqref{eq:Ham-g-APD}, the Hamiltonian $H_g$ for our specific choice of
$g_{ab}$ takes the form
\beq
	H_g= \hbar\omega \big(\sinh(A) K_{+}+ 2\cosh(A) K_0+\sinh(A) K_{-}\big)\ . \label{eq:Ham-su11}
\eeq
Then, using the rearrangement identity \eqref{eq:id1} from Appendix~\ref{app:SU11}, 
we can rewrite the time evolution operator
$e^{-i\frac{H_gt}{\hbar}}$ as
\beq
	e^{-i\frac{H_g t}{\hbar}}= e^{-\beta(t) K_{+}}e^{\ln(\delta(t)) K_0} e^{ -\beta(t)K_{-}}\ ,
\eeq
where $\beta(t),\delta(t)$ are functions of $t,\omega,A$, and are given explicitly in 
Eqs.~\eqref{eq:beta} and \eqref{eq:delta} of Appendix~\ref{app:SU11}.
If we now use the fact that $K_{-}|\psi_0\ran= 0$, then we find that
\beq
	|\psi(t)\ran = e^{-\beta(t) K_{+}}e^{\ln(\delta(t)) K_0} |\psi_0\ran\ .
\eeq
In addition, from the definition of $K_0$ it is clear that $K_0|\psi_0\ran= \frac{\ep_0}{2}|\psi_0\ran$ 
(since $K_0= \frac{1}{2}\frac{H_0}{\hbar\omega}$), and so our final answer for the time-evolved state is
\beq
	|\psi(t)\ran = [\delta(t)]^{\frac{\ep_0}{2}}e^{-\beta(t) K_{+}} |\psi_0\ran\ . \label{eq:post-quench-state}
\eeq
We see that the quench excites all even spin excitations, since acting with $K_{+}$ changes the angular momentum 
of a state by $-2\hbar$ (recall that the angular momentum operator $L_z$ for the CSMM has the form shown in 
Eq.~\eqref{eq:Lz}). Indeed, we can write $K_{+}= \frac{1}{2}\text{Tr}\{(Z^{+})^2\}$ in terms of the 
matrix-valued operator $Z^{+}$, and $\text{Tr}\{(Z^{+})^2\}$ is the operator which creates spin-2 excitations over
the ground state $|\psi_0\ran$ of the original CSMM (recall the form of the excited states $|\{c_1,\dots,c_N\}\ran$
for the original CSMM from Eq.~\eqref{eq:excited-states}). 

We close this subsection by computing the quantity $|\lan\psi_0|\psi(t)\ran|^2$, which is also known as the
\emph{quantum fidelity} or \emph{Loschmidt echo}. The result is
\footnote{Here and in the rest of the paper we assume that 
$|\psi_0\ran$ has been properly normalized.}
\begin{align}
	|\lan\psi_0|\psi(t)\ran|^2 =& \left[\delta(t)\ov{\delta(t)}\right]^{\frac{\ep_0}{2}} \nnb \\
	=& \left[ \cos^2(\omega t) + \cosh^2(A)\sin^2(\omega t)\right]^{-\ep_0} \nnb \\
	=& \left[ 1 + \sinh^2(A)\sin^2(\omega t)\right]^{-\ep_0}\ ,
\end{align}
where we plugged in for $\delta(t)$ using the explicit expression from Appendix~\ref{app:SU11}.
Since $\sin^2(\omega t)$ has a period of $T=\frac{\pi}{\omega}$, we find that the quantum fidelity
oscillates at the period 
\beq
	T= \frac{\pi}{\omega}\equiv \frac{2\pi\hbar}{E_2}\ ,
\eeq
where 
\beq
	E_2:= 2\hbar\omega \label{eq:E2}
\eeq
is the gap for spin-2 excitations in the CSMM. The actual magnitude of the overlap depends on 
the filling fraction $\nu=\frac{1}{m}$ through the power of $\ep_0$. Note also that since 
$\sinh^2(A)\sin^2(\omega t) \geq 0$, the fidelity satisfies $|\lan\psi_0|\psi(t)\ran|^2\leq 1$. 

Recall that the parameter $\ep_0$ has the form $\ep_0= \frac{1}{2}m N^2 + \frac{(1-m)}{2}N$, and
so the quantum fidelity $|\lan\psi_0|\psi(t)\ran|^2$ has a factor of $N^2$ appearing in the exponent. To eliminate this
large factor it is convenient to compare the values of the quantum fidelity between integer and fractional cases. 
Let $\mathcal{F}_{\frac{1}{m}}(t)$ denote the fidelity $|\lan\psi_0|\psi(t)\ran|^2$ for the CSMM with $\theta=\ell_B^2 m$
corresponding to the $\nu=\frac{1}{m}$ Laughlin state. Then we consider the following ratio of $\mathcal{F}_{\frac{1}{m}}(t)$ with
$\mathcal{F}_1(t)$ raised to the $m^{th}$ power,
\beqa
	\frac{\left[\mathcal{F}_1(t)\right]^m}{\mathcal{F}_{\frac{1}{m}}(t)} &=& \left[ 1 + \sinh^2(A)\sin^2(\omega t)\right]^{-\left(\frac{m-1}{2}\right)N} \nnb \\
	&=&  \left[ 1 + \sinh^2(A)\sin^2(\omega t)\right]^{-\varsigma N}\ , \label{eq:fidelity-finite-A}
\eeqa
where 
\beq
	\varsigma=\frac{m-1}{2}
\eeq
is the \emph{anisospin}~\cite{GGB,gromov-son} for the $\nu=\frac{1}{m}$ Laughlin state, 
also called (minus) the \emph{guiding center spin}~\cite{haldane2009,haldane2011}. For small anisotropy 
$A\ll 1$ this is ratio is approximately equal to
\beq
	\frac{\left[\mathcal{F}_1(t)\right]^m}{\mathcal{F}_{\frac{1}{m}}(t)}\approx  1- \varsigma N A^2\sin^2(\omega t)\ .  \label{eq:fidelity-small-A}
\eeq
We see that by comparing the fidelity for $\nu=\frac{1}{m}$ with the fidelity for $\nu=1$, we are able
to extract the universal data $\varsigma$ which characterizes the $\nu=\frac{1}{m}$ Laughlin state. This type of 
comparison with the $\nu=1$ state is very similar to the comparison which is used to extract the dipole moment
per unit length at the edge of a FQH state~\cite{wiegmann2012,park-haldane} 
(the dipole moment also happens to be proportional to the same parameter $\varsigma$ as it is closely related to the
guiding center part of the bulk Hall viscosity).

Finally, for comparison to numerics it is useful to rewrite Eq.~\eqref{eq:fidelity-finite-A} in terms of the filling fraction 
$\nu=\frac{1}{m}$ and the energy gap $E_2=2\hbar\omega$ for the spin-2 mode, which gives 
\begin{subequations}
\label{eq:fidelity-nu}
\begin{align}
	\frac{\left[\mathcal{F}_{1}(t)\right]^{\nu^{-1}}}{\mathcal{F}_{\nu}(t)} &= \left[ 1 + \sinh^2(A)\sin^2\left(\frac{E_2 t}{2\hbar}\right)\right]^{-\varsigma N}\\
	&\approx  1- \varsigma N A^2\sin^2\left(\frac{E_2 t}{2\hbar}\right)\ , \label{eq:Loschmidt-small-A}
\end{align}
\end{subequations}
where in the second line we Taylor-expanded the result for small $A$.
In this form, the expression for $\frac{\left[\mathcal{F}_{1}(t)\right]^{\nu^{-1}}}{\mathcal{F}_{\nu}(t)}$ 
suggests a way to extract the anisospin $\varsigma$ and the spin-2 gap $E_2$ for a general 
FQH state\footnote{Or at least any FQH state which is well-described by projection into a single Landau level.} 
with filling fraction $\nu$ by fitting numerical data from the simulation of a geometric quench for that FQH state to 
Eq.~\eqref{eq:Loschmidt-small-A}.  Indeed, preliminary numerical results~\cite{ZZ-email} suggest that the formula 
\eqref{eq:Loschmidt-small-A} is a good fit to the quantum fidelity for the geometric quench considered in \cite{LGP}.
Note also that for comparison to numerics $E_2$ is expected to equal the energy gap of the
GMP mode at $\mb{k}=0$.

\subsection{Dynamics of the intrinsic metric}

In this subsection we present the exact calculation of the dynamical metric $\hat{g}_{ab}(t)$. We then show that
for small anisotropy $A$, the CSMM result agrees with the bimetric theory results of Ref.~\onlinecite{LGP}. We give a
more detailed comparison with bimetric theory in Sec.~\ref{sec:bimetric}.

We start by using the form of $|\psi(t)\ran$ derived in the last subsection to write the formula for the
inverse metric $\hat{g}^{ab}(t)$ in the form
\beq
	 \hat{g}^{ab}(t)= \frac{2}{\ep_0} \left[\delta(t)\ov{\delta(t)}\right]^{\frac{\ep_0}{2}}\lan \psi_0 |e^{-\ov{\beta(t)} K_{-}}\mathsf{\Lambda}^{ab}e^{-\beta(t) K_{+}}|\psi_0\ran\ .
\eeq
We know that the $sl(2,\mathbb{R})$ generators $\mathsf{\Lambda}^{ab}$ can be expressed in terms of the 
$su(1,1)$ generators $K_0,K_{\pm}$, and so we choose to proceed with this calculation by first calculating the expectation 
values $\lan \psi_0 |e^{-\ov{\beta(t)} K_{-}}K_0 e^{-\beta(t) K_{+}}|\psi_0\ran$ and 
$\lan \psi_0 |e^{-\ov{\beta(t)} K_{-}}K_{\pm} e^{-\beta(t) K_{+}}|\psi_0\ran$. 

To calculate these expectation values we use a generating function technique. We define a function $f(a,b,c)$ of three
variables $a,b,c$ by
\beq
	f(a,b,c):= \lan \psi_0| e^{a K_{-}}e^{b K_0} e^{c K_{+}} |\psi_0\ran\  .
\eeq
Then the expectation values which we are interested in can be computed from $f(a,b,c)$ as
\begin{widetext}
\begin{align}
	\lan \psi_0 |e^{-\ov{\beta(t)} K_{-}}K_{-} e^{-\beta(t) K_{+}}|\psi_0\ran &= \frac{\pd f(a,b,c)}{\pd a}\Bigg|_{a=-\ov{\beta(t)},\ b=0,\ c=-\beta(t)} \\
	\lan \psi_0 |e^{-\ov{\beta(t)} K_{-}}K_0 e^{-\beta(t) K_{+}}|\psi_0\ran &= \frac{\pd f(a,b,c)}{\pd b}\Bigg|_{a=-\ov{\beta(t)},\ b=0,\ c=-\beta(t)} \\
	\lan \psi_0 |e^{-\ov{\beta(t)} K_{-}}K_{+} e^{-\beta(t) K_{+}}|\psi_0\ran &= \frac{\pd f(a,b,c)}{\pd c}\Bigg|_{a=-\ov{\beta(t)},\ b=0,\ c=-\beta(t)} \ .
\end{align}
\end{widetext}
The function $f(a,b,c)$ itself can be calculated using the rearrangement identity Eq.~\eqref{eq:id2} from 
Appendix~\ref{app:SU11}, combined with the fact that $K_{-}|\psi_0\ran=0$. Using that information, we 
find that 
\beq
	f(a,b,c)= [b'(a,b,c)]^{\frac{\ep_0}{2}}\ ,
\eeq
where the new function $b'(a,b,c)$ is written down explicitly in Eq.~\eqref{eq:b-prime} of Appendix~\ref{app:SU11}.

\begin{widetext}
The calculation now proceeds in a straightforward manner and we find that the metric $\hat{g}_{ab}(t)$ (which is the inverse of
$\hat{g}^{ab}(t)$) can be written in matrix form as
\beq
	\hat{g}(t)=\frac{1}{1-|\beta(t)|^2} \begin{pmatrix}
	(1+\beta(t))(1+\ov{\beta(t)}) & -i (\beta(t)-\ov{\beta(t)}) \\
	-i (\beta(t)-\ov{\beta(t)})  & (1-\beta(t))(1-\ov{\beta(t)})
\end{pmatrix}\ .
\eeq
where $\beta(t)$ is again the function defined in Eq.~\eqref{eq:beta} of Appendix~\ref{app:SU11}. We also note here that in order to derive 
these expressions we needed to use the formula Eq.~\eqref{eq:id1-extra}. 
\end{widetext}

%We see that the dynamical metric has the same form as the parametrization for a unimodular metric from
%Eq.~\eqref{eq:unimodular-metric}, with the parameter $\beta(t)$ given by Eq.~\eqref{eq:beta} from 
%Appendix~\ref{app:SU11}, which we reproduce here for convenience,
%\beq
%	\beta(t) = \frac{\sinh(A)}{\cosh(A)-i\cot(\omega t)}\ .
%\eeq
%We see that it depends on $A$ (which defines the metric $g_{ab}$ we used for the quench) and also on $\omega$ and $t$.

From Eq.~\eqref{eq:beta} we can see that the parameter $\beta(t)$ oscillates with a period of 
$T=\frac{\pi}{\omega}=\frac{2\pi\hbar}{E_2}$ ($E_2=2\hbar\omega$ was defined in Eq.~\eqref{eq:E2}), 
and its time average is 
\beq
	\lan \beta(t) \ran = \frac{1}{T}\int_0^T dt\ \beta(t)= \tanh\left(\frac{A}{2}\right)\ .
\eeq
It is interesting to note that if we replace $\beta(t)$ with $\lan \beta(t) \ran$ in the metric $\hat{g}_{ab}(t)$, 
then the dynamical metric reduces to the metric $g_{ab}$ from Eq.~\eqref{eq:anisotropy-g} 
that we used for the quench Hamiltonian $H_g$.

We now study the CSMM solution for the dynamical metric in the case of small anisotropy $A\ll 1$, because in this case we can
compare to the results of \cite{LGP} obtained using the linearized equations of motion of bimetric theory. To compare our 
dynamical metric $\hat{g}_{ab}(t)$ with the one obtained in \cite{LGP}, we write the complex parameter $\beta(t)$ in terms 
of a real parameter $Q(t)$ and a real phase $\phi(t)$ as 
\beq
	\beta(t)= \tanh\left(\frac{Q(t)}{2}\right)e^{i\phi(t)}\ .
\eeq
With this parametrization the dynamical metric $\hat{g}_{ab}(t)$ takes the form shown in Eq.~\eqref{eq:LGP-metric} 
and used in \cite{LGP}. {Note that in this parametrization one should always choose $Q(t)\geq 0$ so
that there is no redundancy in the description (all information about the phase of $\beta(t)$ should be packaged
in the parameter $\phi(t)$). In this case we find that $Q(t)$ is related to $\beta(t)$ as
\beq
	Q(t)= 2\tanh^{-1}\left[\sqrt{|\beta(t)|^2}\right]\ ,\quad \phi(t)= \text{arg}[\beta(t)]\ .
\eeq

For small anisotropy the parameter $Q(t)$ in the solution for the dynamical metric is expected to be small, and so
in this case we can write
\beq
	\beta(t)\approx \frac{Q(t)}{2}e^{i\phi(t)}\ .
\eeq
On the other hand, for small $A$ the exact solution for $\beta(t)$ from the CSMM takes the form
\beqa
	\beta(t) &\approx& \frac{A}{1-i\cot(\omega t)} \nnb \\
	&=& A\sin(\omega t)e^{-i\omega t+i\frac{\pi}{2}}\ .
\eeqa
By comparing these two expressions for $\beta(t)$ we obtain the solution for $Q(t)$ and $\phi(t)$ for the case of small
$A$ (we assume $A>0$),
\beqa
	Q(t) &=& 2 A \left|\sin\left(\frac{E_2}{2\hbar}t\right)\right| \\
	\phi(t) &=& \pi - \frac{E_2}{2\hbar}t - \frac{\pi}{2}\text{sgn}\left[\sin\left(\frac{E_2}{2\hbar}t\right)\right]\ ,
\eeqa
where $E_2=2\hbar\omega$ is the gap for the spin-2 mode in the CSMM. These equations exactly match the predictions of
bimetric theory, as these solutions are identical to Eq.~5 of \cite{LGP}, which is a solution to the linearized equations of
motion Eqs.~20 and 21 of \cite{LGP} for the geometric quench in bimetric theory\footnote{
For comparison to \cite{LGP} note that $\hbar=1$ in that paper. Also, here we use the convention that
$Q(t)\geq 0$ to avoid redundancy in the parametrization of $\beta(t)$ in terms of $Q(t)$ and $\phi(t)$. This explains
the slight difference between our linearized solution and Eq.~5 of \cite{LGP}.}. 

For the case of arbitrary anisotropy $A$ the CSMM predicts that the dynamical metric 
$\hat{g}_{ab}(t)$ undergoes \emph{nonlinear} oscillations, in the sense that the amplitude of the function $\beta(t)$ is a 
nonlinear function of $A$. However, these oscillations still have a definite period $T=\frac{2\pi \hbar}{E_2}$ 
set by the energy gap $E_2$ of the spin-2 mode in the CSMM, so the period of the oscillations is independent of the
amplitude.

For small anisotropy $A\ll 1$ the linearized solution and the full solution are nearly identical. However,
the difference between these two solutions can be seen clearly in the case of a quench with large anisotropy (see Fig.~\ref{fig:quench-plot} for details).

\begin{figure}[t]
  \centering
    \includegraphics[width= .4\textwidth]{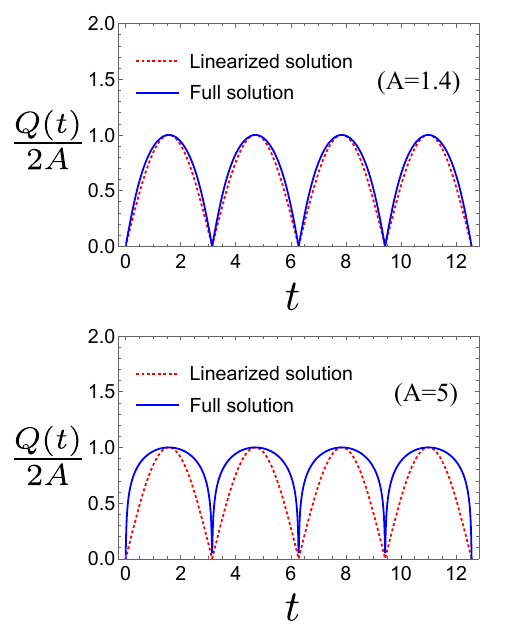} 
\vskip 5pt
 \caption{Plots of the linearized solution for $Q(t)$ (dotted red line) and the full nonlinear solution (blue line) for
the case of medium anisotropy $A=1.4$ and large anisotropy $A=5$, both normalized by dividing by
$2A$. The time axis is plotted in units with $\omega=1$. The maximum value of $Q(t)/(2A)$ is $1$ and occurs (for 
$\omega=1$) when $t= \pi/2 + \pi n$ for any integer $n$. For smaller $A$ the linearized solution is close to the full solution, 
while for larger $A$ the full nonlinear solution has a much rounder profile than the linearized solution.}
\label{fig:quench-plot}
\end{figure}

\subsection{Extension to the non-Abelian Blok-Wen states}

We close this section with a discussion of how our results extend to the matrix model for the non-Abelian Blok-Wen series of FQH 
states~\cite{tong2016,dorey2016matrix} (see also Ref.~\onlinecite{LHTT} for the calculation of the Hall viscosity in this matrix model).

The main difference between the CSMM and the non-Abelian matrix model (NAMM) is that instead of having just one
complex vector $\vphi$, the NAMM has $p$ complex vectors $\vphi_{\al}$, $\al=1,\dots,p$, for some positive integer $p$.
The action for the NAMM takes the form
\begin{align}
	S^{\text{{\tiny{(NA)}}}}_0 &= -\frac{eB}{2}\int_0^T dt\ \text{Tr}\Big\{ \ep_{ab}X^a \D_0 X^b + 2\theta A_0 \nnb \\
+&\ \omega \delta_{ab}X^a X^b \Big\} + i \sum_{\al=1}^p \int_0^T dt\ \ov{\vphi}^{T}_{\al}\D_0 \vphi_{\al}\ ,
\end{align}
and one can see that this action has an additional $SU(p)$ global symmetry which rotates the different $\vphi_{\al}$ into
each other. In this model it is also convenient to parametrize $\theta$
(which is still quantized to be an integer) as
\beq
	\theta= \ell_B^2 (k+p)\ ,
\eeq	
for some other integer $k$, and we assume that $k$ is chosen so that $k+p\geq 0$. The NAMM then describes the
subset of the Blok-Wen states at filling
\beq
	\nu= \frac{p}{k+p}\ ,
\eeq
and the $\nu=1/(k+1)$ Laughlin state is recovered from this model upon setting $p=1$ (so for $p=1$, set $k+1=m$ to compare 
with our previous results on the CSMM). In addition, it is known~\cite{LHTT} that the anisospin $\varsigma$ for these states is 
independent of $p$ and given by
\beq
	\varsigma=\frac{k}{2}\ .
\eeq

We now give a brief summary of the quantization of this model. First,
the $b_j$ variables from earlier acquire an additional index $\al$, so that we now have $p N$ oscillator variables
$b^j_{\al}$ and their Hermitian conjugates $b^{\dg}_{\al,j}$, and these obey 
$[b^j_{\al},b^{\dg}_{\beta,\ell}]= \delta^j_{\ell} \delta_{\al\beta}$. Next, the constraint enforced by 
$A_0$ is now modified to
\beq
	G := 	ieB[X^1,X^2]_M+eB\theta\mathbb{I} - \sum_{\al=1}^p\vphi_{\al}\ov{\vphi}^T_{\al}= 0 \ . \label{eq:constraint-NA}
\eeq
The $SU(N)$ part of this constraint still requires physical states to be $SU(N)$ singlets, but the $U(1)$ part of the constraint
now takes the form
\beq
	\sum_{\al=1}^p b^{\dg}_{\al,j} b^j_{\al}|\text{phys}\ran= Nk |\text{phys}\ran \label{eq:constraint-u1-NA}
\eeq	
for all physical states $|\text{phys}\ran$. On the other hand, the Hamiltonian $H_0$ (in the $A_0=0$ gauge) and angular 
momentum operator $L_z$ for the NAMM are identical to those in the CSMM. Thus, anisotropy parametrized by 
$g_{ab}$ is introduced into the Hamiltonian in the same way as for the CSMM and the
geometric quench protocol for the NAMM is exactly the same as for the CSMM. 

Finally, we come to the construction of the quantum ground state of the NAMM. Here we consider only the case where
$N$ is divisible by $p$, because in this case the ground state is unique (see Ref.~\onlinecite{tong2016,dorey2016matrix} for more 
details and the general case). To construct the ground state we first construct, for any integer $r\geq 0$, the operator
\beq
 \mathcal{B}^{\dg}(r)_{j_1\cdots j_p} := \ep^{\al_1\cdots\al_p}[b^{\dg}_{\al_1}(Z^{\dg})^r]_{j_1}\cdots[b^{\dg}_{\al_p}(Z^{\dg})^r]_{j_p}\ .
\eeq
This operator is a singlet under the global $SU(p)$ symmetry of the model, but it is not invariant under the $SU(N)$
gauge symmetry. An operator which is invariant under both the $SU(p)$ global symmetry and the $SU(N)$ gauge symmetry can 
then be constructed from the $\mathcal{B}^{\dg}(r)_{j_1\cdots j_p}$ operators as
\begin{align}
\tilde{\mathcal{B}} := \ep^{j_1\cdots j_N}&\mathcal{B}^{\dg}(0)_{j_1\cdots j_p}\mathcal{B}^{\dg}(1)_{j_{p+1}\cdots j_{2p}} \nnb \\
&\cdots\mathcal{B}^{\dg}({N}/{p}-1)_{j_{N-p+1}\cdots j_N}\ .
\end{align}
Finally, the unique ground state of the NAMM can be constructed using $\tilde{\mathcal{B}}$ as
\beq
	|\psi^{\text{{\tiny{(NA)}}}}_0\ran = \tilde{\mathcal{B}}^k|0\ran \ .
\eeq
In particular, the power of $k$ here ensures that Eq.~\eqref{eq:constraint-u1-NA} is satisfied. The energy of the 
ground state is 
\beq
	E^{\text{{\tiny{(NA)}}}}_0= \hbar\omega\left[ \left(\frac{k+p}{p}\right)\frac{N^2}{2} -\frac{k}{2}N\right]\ ,
\eeq
and we again define the dimensionless quantity
\beq
	\ep^{\text{{\tiny{(NA)}}}}_0 := \frac{E^{\text{{\tiny{(NA)}}}}_0}{\hbar\omega}\ .
\eeq

We are now ready to explain how our results generalize to the NAMM. The key point is that one can still construct the
$su(1,1)$ generators $K_0,K_{\pm}$ as before and, crucially, we still have the property that $K_{-}|\psi_0\ran=0$. The
proof of this fact is exactly the same as the proof we gave in the CSMM case. This fact implies that our results for
the geometric quench in the CSMM carry over to the NAMM with the trivial replacement 
$\ep_0 \to \ep^{\text{{\tiny{(NA)}}}}_0$ in all formulas.
For the post-quench state in the NAMM we find
\beq
	|\psi^{\text{{\tiny{(NA)}}}}(t)\ran = [\delta(t)]^{\frac{\ep^{\text{{\tiny{(NA)}}}}_0}{2}}e^{-\beta(t) K_{+}} 
	|\psi^{\text{{\tiny{(NA)}}}}_0\ran\ .
\eeq
Here we emphasize that even though the NAMM has an excitation spectrum which is much more complicated than the 
CSMM, the geometric quench still only excites the spin-2 excitations which are created by $K_{+}$. For the quantum fidelity
we find (again, assuming that $|\psi^{\text{{\tiny{(NA)}}}}_0\ran$ has been properly normalized)
\beq
	|\lan\psi^{\text{{\tiny{(NA)}}}}_0|\psi^{\text{{\tiny{(NA)}}}}(t)\ran|^2 	= \left[ 1 + \sinh^2(A)\sin^2(\omega t)\right]^{-\ep^{\text{{\tiny{(NA)}}}}_0}\ .
\eeq
In particular, Eqs.~\eqref{eq:fidelity-nu} still hold in this case, 
with the appropriate values $\nu=\frac{p}{k+p}$ and $\varsigma= \frac{k}{2}$ for the Blok-Wen states. Finally, we define the 
dynamical metric in the NAMM as (compare to Eq.~\eqref{eq:metric-time-dependent})
\beq
	\hat{g}^{ab}(t):= \frac{2}{\ep^{\text{{\tiny{(NA)}}}}_0}\lan \psi^{\text{{\tiny{(NA)}}}}(t)|\mathsf{\Lambda}^{ab}|\psi^{\text{{\tiny{(NA)}}}}(t)\ran\ ,
\eeq
and with this definition we find that $\hat{g}_{ab}(t)$ for the NAMM is identical to the answer found for the CSMM. 

We conclude that the geometric quench excites the same dynamics in the Laughlin and Blok-Wen states, despite the fundamentally different topological order. Indeed, 
 both states support the gapped GMP mode, and the geometric quench excites the same dynamics for this mode in both sets of states.

\section{Comparison with bimetric theory}
\label{sec:bimetric}

In this section we present a more detailed comparison between the geometric quench in the CSMM\footnote{For brevity,
in the rest of the paper we mostly refer to our results on the CSMM. However, the reader should keep in mind that we have
demonstrated in the previous section that our results on the geometric quench in the CSMM also apply to the
non-Abelian matrix model for the Blok-Wen states.} 
and in bimetric theory.
We derive the differential equation obeyed by the dynamical metric $\hat{g}_{ab}(t)$ in the CSMM, and we show that
this differential equation is not an exact match to the differential equations obtained within bimetric theory in 
\cite{LGP}. We then suggest an alternative (and simpler) potential energy term for the bimetric theory Lagrangian, and
we show that the equations of motion for bimetric theory with this simpler potential energy exactly match
the equations of motion for $\hat{g}_{ab}(t)$ in the CSMM.

\subsection{Differential equations obeyed by the dynamical metric}

To derive the differential equation obeyed by the dynamical metric in the CSMM we return to the relation
\beq
	\hat{g}^{ab}(t) = \frac{2}{\ep_0}\lan \psi(t)|\mathsf{\Lambda}^{ab}|\psi(t)\ran
\eeq
and differentiate with respect to time,
\beqa
	\dot{\hat{g}}^{ab}(t) &=& \frac{2}{\ep_0}\frac{i}{\hbar}\lan \psi(t)|[H_g ,\mathsf{\Lambda}^{ab}]|\psi(t)\ran \nnb \\
	&=& -i\frac{2\omega}{\ep_0}g_{cd} \lan \psi(t)|[\mathsf{\Lambda}^{ab},\mathsf{\Lambda}^{cd}]|\psi(t)\ran\ ,
\eeqa
where we used $H_g= \hbar\omega g_{ab}\mathsf{\Lambda}^{ab}$.
Next, we use the commutation relations of the $sl(2,\mathbb{R})$ generators (Eq.~\eqref{eq:SL2R}
to find that $\hat{g}^{ab}(t)$ obeys the \emph{linear} differential equation
\beq
	\dot{\hat{g}}^{ab}(t)= \frac{\omega}{2}g_{cd}\left[ \ep^{bc}\hat{g}^{ad}(t)+\ep^{bd}\hat{g}^{ac}(t) + \ep^{ac}\hat{g}^{bd}(t)+\ep^{ad}\hat{g}^{bc}(t)         \right]\ . \label{eq:metric-ODE}
\eeq

We now choose the anisotropy metric $g_{ab}$ as in Eq.~\eqref{eq:anisotropy-g} and we parametrize $\hat{g}_{ab}(t)$
in terms of two variables $Q(t)$ and $\phi(t)$ as in Eq.~\eqref{eq:LGP-metric}. This means that the inverse metric 
$\hat{g}^{-1}(t)$ takes the form
\begin{widetext}
\beq
	\hat{g}^{-1}(t) =\begin{pmatrix}
	\cosh(Q)-\cos(\phi)\sinh(Q) & -\sin(\phi)\sinh(Q) \\
	-\sin(\phi)\sinh(Q)  & \cosh(Q)+\cos(\phi)\sinh(Q)
\end{pmatrix}\ .
\eeq	
\end{widetext}
In this case the linear differential equation \eqref{eq:metric-ODE} for $\hat{g}^{ab}(t)$ reduces to two coupled 
nonlinear differential equations for $\phi$ and $Q$,
\begin{subequations}
\label{eq:CSMM-metric-eqns}
\begin{align}
	\dot{Q}&= 2\omega \sinh(A)\sin(\phi) \\
	\dot{\phi}\sinh(Q) &= \nnb\\
2\omega &\Big( \sinh(A)\cos(\phi)\cosh(Q) - \cosh(A)\sinh(Q) \Big)\ .
\end{align}
\end{subequations}

\subsection{Comparison with bimetric equations}

We now compare Eqs.~\eqref{eq:CSMM-metric-eqns} to the predictions of bimetric theory. Here we briefly
recall the form of the Lagrangian for bimetric theory (as considered in the quench calculation of \cite{LGP}). For more
details on bimetric theory we refer the reader to Refs.~\onlinecite{GGB,gromov-son}.

The degree of freedom in bimetric theory is a dynamical unimodular metric, which we denote here by $\hat{g}_{ab}(\mb{x},t)$, where
$\mb{x}=(x^1,x^2)$ are coordinates on two-dimensional space, and $t$ is the 
time\footnote{Refs.~\onlinecite{GGB,gromov-son,LGP} use
$i,j,k,...$ for spatial indices and $a,b,c,...$ for internal $SO(2)$ indices on frame and coframe fields. Here we depart from
their convention and use $a,b=1,2$ for spatial indices in order to match our conventions for the CSMM. No confusion should 
arise as our discussion here does not require the introduction of frame or coframe fields.}. 
Physically, the field $\hat{g}_{ab}(\mb{x},t)$ corresponds to the gapped spin-2 mode which is equal to the long-wavelength 
(small $\mb{k}$) limit of the gapped GMP mode~\cite{GMP} (recall that the GMP mode has a definite angular momentum equal to 
$2\hbar$ near $\mb{k}=0$). Note also that because of the constraint that $\hat{g}_{ab}(\mb{x},t)$ is a \emph{unimodular} 
metric (i.e., it has determinant equal to one), the bimetric theory of Refs.~\onlinecite{GGB,gromov-son} does not contain a spin-0 ``dilaton'' 
mode.

In the specific case of the geometric quench problem, in which anisotropy is represented by the constant metric 
$g_{ab}$ of Eq.~\eqref{eq:anisotropy-g}, the dynamical metric in bimetric theory can be taken to be independent of
space, $\hat{g}_{ab}(\mb{x},t)\to \hat{g}_{ab}(t)$, and the Lagrangian of bimetric theory consists of two terms 
\beq
	\mathscr{L}=\mathscr{L}_{top}+\mathscr{L}_{pot}\ .
\eeq
The first term $\mathscr{L}_{top}$ is the topological term in the bimetric theory Lagrangian, and it has the form (here we 
assume a parametrization of $\hat{g}_{ab}(t)$ as in Eq.~\eqref{eq:LGP-metric})
\beq
	\mathscr{L}_{top}= \frac{\varsigma\ov{\rho}}{2}(1-\cosh(Q))\dot{\phi}\ ,
\eeq
where $\varsigma$ is the \emph{anisospin} of the FQH state and $\ov{\rho}=\frac{\nu}{2\pi\ell_B^2}$ is the mean particle density
of the state. The potential energy term incorporates the anisotropy metric $g_{ab}$ and takes the form
\begin{align}
	\mathscr{L}_{pot} = -\frac{m}{2}\left[\frac{1}{2}g^{ab}\hat{g}_{ab}-\gamma\right]^2
\end{align}
where $m>0$ and $\gamma$ are parameters appearing in the bimetric theory. In particular, the parameter $\gamma$ 
allows for the possibility to realize the nematic quantum Hall transition within bimetric theory, and this transition occurs
at $\gamma=1$ (the gapped FQH phase corresponds to $\gamma<1$).

The differential equations for the geometric quench in bimetric theory, which were obtained in \cite{LGP} by varying the 
Lagrangian $\mathscr{L}=\mathscr{L}_{top}+\mathscr{L}_{pot}$, take the form (Eqs.~15 and 16 of \cite{LGP})
\begin{widetext} 
\beq
	\dot{Q}= 2\Omega \sinh(A)\sin(\phi) \Big( -\gamma-\sinh(A)\sinh(Q)\cos(\phi)+\cosh(A)\cosh(Q) \Big)
\eeq
and
\beqa
	\dot{\phi}\sinh(Q) &=& 2\Omega \Big( \sinh(A)\cos(\phi)\cosh(Q) - \cosh(A)\sinh(Q) \Big)\nnb \\
&\times& \Big( -\gamma-\sinh(A)\sinh(Q)\cos(\phi)+\cosh(A)\cosh(Q) \Big)\ ,
\eeqa
where $\Omega= \frac{m}{\ov{\rho}\varsigma}$.
The only difference between these equations and Eqs.~\eqref{eq:CSMM-metric-eqns} for the quench in the CSMM is that the 
constant factor of $2\omega$ in Eqs.~\eqref{eq:CSMM-metric-eqns} is replaced by the large factor
\beq
	2\Omega\Big( -\gamma-\sinh(A)\sinh(Q)\cos(\phi)+\cosh(A)\cosh(Q) \big) = 2\Omega\left( 
\frac{1}{2}g^{ab}\hat{g}_{ab}-\gamma\right)\ ,
\eeq
\end{widetext}
which has explicit dependence on the dynamical fields $Q(t)$ and $\phi(t)$ which parametrize $\hat{g}_{ab}(t)$.

For small anisotropy (small $A$ and, hence, small $Q$), we have
\beq
	2\Omega\left( \frac{1}{2}\hat{g}_{ab} g^{ab}-\gamma\right) \to 2\Omega(1-\gamma)\ ,
\eeq
and
\beq
	E_{\gamma}:= 2\Omega(1-\gamma)
\eeq
is interpreted in bimetric theory as the gap of the spin-2 mode at $\mb{k}=0$. On the other hand,
we know that $E_2= 2\omega$ (we set $\hbar=1$ here to compare with Ref.~\onlinecite{LGP}) is the gap for the spin-2 excitation 
in the CSMM. Thus, it appears that
while the CSMM has a constant gap of $2\omega$ for the spin-2 mode, the bimetric theory with potential 
$\mathscr{L}_{pot}$ can be interpreted as having a \emph{field-dependent} gap 
$2\Omega\left( \frac{1}{2}g^{ab}\hat{g}_{ab} -\gamma\right)$, and this field-dependent gap only reduces to a constant in 
the regime of small anisotropy and small fluctuations of the dynamical metric. 
This field-dependent gap can be thought of as arising from the nontrivial interaction in bimetric
theory with the potential $\mathscr{L}_{pot}$, which is \emph{quadratic} in the dynamical metric and, therefore,
\emph{quartic} in the coframe field which is the true degree of freedom in bimetric theory.

These findings suggest that the main difference between the predictions of the CSMM and of bimetric theory stems from the
particular choice of potential energy term $\mathscr{L}_{pot}$ for bimetric theory. This raises the question of whether 
there exists a different choice of potential energy term, say $\mathscr{L}'_{pot}$, such that the equations of motion in
the bimetric theory with this new potential energy term coincide with the equations derived from the CSMM. We 
construct such a potential energy term in the next subsection.

\subsection{A new potential energy term for bimetric theory, and an exact match with the CSMM}

In this subsection we show that the differential equations for the intrinsic metric $\hat{g}_{ab}(t)$ derived in the CSMM 
can be reproduced by a variant of the bimetric theory which features a different potential energy term than the one used in
\cite{LGP}. The modified potential energy term that we consider has the form
\beq
	\mathscr{L}'_{pot}= -\frac{m'}{2}g^{ab}\hat{g}_{ab}\ ,
\eeq
where $m'>0$ is a new phenomenological parameter with units of (length)$^{-2}$(time)$^{-1}$.
This term is chosen to mimic the 
form of the Hamiltonian $H_g=\hbar\omega g_{ab}\mathsf{\Lambda}^{ab}$ in the CSMM. The main difference between 
$\mathscr{L}_{pot}$ and $\mathscr{L}'_{pot}$ is that the latter allows for a single, isotropic phase, whereas the former 
supports two phases: isotropic and (gapless) nematic phase, which spontaneously breaks rotational symmetry.
In terms of $Q$, $\phi$, and $A$ this term takes the form
\beq
	\mathscr{L}'_{pot}= m' \Big( \sinh(A)\sinh(Q)\cos(\phi)-\cosh(A)\cosh(Q) \Big)\ .
\eeq

The equations of motion for the modified bimetric theory with Lagrangian
\beq
	\mathscr{L}'=\mathscr{L}_{top}+\mathscr{L}'_{pot}
\eeq
exactly match the CSMM equations \eqref{eq:CSMM-metric-eqns} if the parameters of bimetric theory are related to
the parameter $\omega$ in the CSMM as
\beq
	\omega = \frac{m'}{\ov{\rho}\varsigma}\ .
\eeq
Therefore we find that there exists an alternative potential energy function for the bimetric theory such that the
bimetric theory and the CSMM give identical answers for the dynamics of the metric $\hat{g}_{ab}(t)$ after a
geometric quench.

Finally, we emphasize that the main qualitative difference between the two potentials considered here is that $\mathscr L^\prime_{pot}$ 
does not support a nematic transition. This has to be the case since the CSMM describes only the gapped quantum Hall phase. Implementing the nematic transition within the CSMM is presently an open problem.

%%%%%
\section{Higher-spin collective modes}
\label{sec:higher-spin}
%%%%%
%edited

In this section we show that in addition to the spin-$2$ collective mode $\hat{g}^{ab}(t)$ in the CSMM, it is possible to introduce an infinite 
tower of higher-spin collective modes, $\hat{g}^{abcd\ldots}(t)$. We then show that the higher-spin modes with even spin are excited by the 
geometric quench and undergo oscillations at frequencies determined by their gap. Higher-spin modes in the FQH effect have quite a long 
history \cite{GMP,CTZ,susskind,P3,HVR,read1998lowest, golkar2016higher, nguyen2018fractional, cappelli2016multipole,LGP}. 
Despite previous theoretical efforts, the dynamics of these modes is still not well understood.

\subsection{Dynamics of the higher-spin modes}

The higher spin collective modes are introduced by generalizing the $K_{\pm}$ and $K_0$ operators studied in the previous sections. 
Specifically, we will consider the single trace operators
\beq
K^{\al_1\al_2 \al_3 \al_4 \cdots} = \text{Tr}\{Z^{\al_1}Z^{\al_2}Z^{\al_3}Z^{\al_4}\cdots\}\,,   
\eeq
where each $\al_j=\pm$. To connect these operators with $K_{\pm}$ and $K_0$ we simply note that 
\begin{subequations}
\beqa
	K_{\pm} &=& \frac{1}{2}K^{\pm\pm}\ , \\
		K_0 &=& \frac{1}{2}K^{+-} + \frac{N^2}{4}\ .
\eeqa
\end{subequations}
The extra constant factor in the relation between $K_0$ and $K^{+-}=K^{-+}$ is not important since $K_0$ and
 $\frac{1}{2}K^{+-}$ still have identical commutators with any other operator. The total spin of the operator $K^{\al_1\al_2 \al_3 \al_4 \cdots}$ is given by the number of indices equal to ``$+$'' 
minus the number of indices equal to``$-$''. More concretely, acting with $K^{\al_1\al_2 \al_3 \al_4 \cdots}$ on a state will change the
angular momentum of that state by $-\hbar\sum_j \al_j$. In particular, it is clear that operators with greater than two indices can have
spin higher than 2.

For every state $|\chi\rangle$ in the Hilbert space of the CSMM we can define intrinsic higher-spin collective variables 
according to
\beq
\hat g^{\al_1\al_2 \al_3 \al_4 \cdots}_{\chi} = \langle \chi| K^{\al_1\al_2 \al_3 \al_4 \cdots} |\chi\rangle\,.
\eeq
Our objective is to quantify the dynamics of $\hat g^{\al_1\al_2 \al_3 \al_4 \cdots}_{\chi}(t)$, for a particular choice of $|\chi\rangle$, namely, the quenched state $|\psi(t)\rangle = e^{-i\frac{H_g t}{\hbar}}|\psi_0\ran$. We will assume that the quenched Hamiltonian is given by \eqref{eq:Ham-su11}\footnote{In principle, we could have studied more complicated Hamiltonians which depend on higher-spin
operators as well as the spin-$2$ operators. However, such Hamiltonians appear to lead to very complicated dynamics which is beyond the scope of the present paper.}. It turns out that finding $\hat g^{\al_1\al_2 \al_3 \al_4 \cdots}(t)$ is already quite a formidable task because the operators $K^{\al_1\al_2 \al_3 \al_4 \cdots}$ do not possess simple commutation relations with each other, with the exception of the spin-$2$ 
$sl(2,\mathbb R)$ subalgebra formed by $\{K^{++}, K^{--}, K^{+-}\}$. It is believed that when properly defined the
operators $K^{\al_1\al_2 \al_3 \al_4 \cdots}$ should obey a $W_\infty$ algebra. Identifying the ``right'' basis in the set of $K^{\al_1\al_2 \al_3 \al_4 \cdots}$ that leads to the $W_\infty$ algebra is an unsolved problem \cite{cappelli2005}. We give some further discussion of this
issue in Appendix~\ref{app:W}.

Given these complications, we limit our considerations in this section to the spin-$4$ collective variable 
\beq
	\hat{g}^{\al_1\al_2\al_3\al_4}(t) := \lan\psi(t)| K^{\al_1\al_2 \al_3 \al_4}|\psi(t)\ran\ ,
\eeq
where $|\psi(t)\rangle = e^{-i\frac{H_g t}{\hbar}}|\psi_0\ran$ is the post-quench state.
The equation of motion for $\hat{g}^{\al_1\al_2\al_3\al_4}(t)$ takes the form
\beq
	\dot{\hat{g}}^{\al_1\al_2\al_3\al_4}(t) = \frac{i}{\hbar}\lan\psi(t)|[H_g, K^{\al_1\al_2\al_3\al_4}]|\psi(t)\ran .
\eeq
To evaluate the right-hand side of this equation, recall that the quench Hamiltonian $H_g$ can be written in terms of the 
$su(1,1)$ generators as in Eq.~\eqref{eq:Ham-su11}.
Then we can evaluate the commutators $[H_g, K^{\al_1\al_2\al_3\al_4}]$ using the following
commutation relations,
\beqa
	\left[K_0, {(Z^{\pm})^j}_k\right] &=& \pm \frac{1}{2}{(Z^{\pm})^j}_k \\
	\left[K_{-},{(Z^{+})^j}_k \right] &=& {(Z^{-})^j}_k \\
	\left[K_{+},{(Z^{-})^j}_k \right] &=& -{(Z^{+})^j}_k\ ,
\eeqa
which are easily derived from the commutation relations of $Z^{\pm}$ and the definition of the $su(1,1)$ generators.
These commutation relations make it clear that the Hamiltonian 
$H_g$ mixes the 16 operators $K^{\al_1\al_2\al_3\al_4}$ among themselves, but \emph{does not} mix 
them with any other operators. This is because taking the commutator of $ {(Z^{\pm})^j}_k$ with any of the $su(1,1)$
generators does not have any effect on the $U(N)$ indices $j$ and $k$.

The resulting evolution equations for the 16 variables $\hat{g}^{\al_1\al_2\al_3\al_4}(t)$ can
be written in a matrix form. To write down this equation we first define a 16-dimensional vector whose components 
$V^J(t)$, $J=1,\dots,16$, are defined in Eq.~\ref{eq:V} of Appendix~\ref{app:big-formulas}. We also define a
$16\times 16$ matrix $M$, which is displayed in Eq.~\ref{eq:M} of Appendix~\ref{app:big-formulas}. Using $V(t)$ and
$M$, the evolution equations for the 16 variables $\hat{g}^{\al_1\al_2\al_3\al_4}(t)$  can be written in the 
concise form
\beq
	\dot{V}(t)= i\omega M V(t)\ .
\eeq

Let us pause here to discuss some properties of the matrix $M$. This matrix is too big to be manipulated by hand,
but it can be handled using Mathematica~\cite{math}. We find that $M$ has eigenvalues
$\pm 4$ with multiplicity one for each sign, $\pm 2$ with multiplicity four for each sign, and $0$ with multiplicity six. In 
addition, one can show that $M$ has sixteen linearly independent eigenvectors\footnote{Mathematica's ``Eigenvectors"
command yields 16 eigenvectors for this matrix which are clearly not orthonormal. However, one can check that these
eigenvectors are linearly independent by studying the determinant of the matrix whose rows are these eigenvectors. We
have checked that this determinant is non-zero for any value of $A$, and so $M$ really does have a full set of 16 linearly
independent eigenvectors.}. It seems, however, that the eigenvectors of $M$ cannot be chosen to be orthogonal while
still remaining eigenvectors of $M$.

The fact that $M$ possesses a set of 16 linearly independent eigenvectors means that we can decompose $M$ as
\beq
	M= S D S^{-1}\ ,
\eeq
where $D$ is a diagonal matrix whose entries are the eigenvalues of $M$ and $S$ is an invertible (but in general not 
orthogonal) matrix whose columns are the eigenvectors of $M$. We can use this decomposition to solve the differential equation by 
defining a new vector 
\beq
	W(t)= S^{-1} V(t)\ .
\eeq
Then one can show that $\dot{W}(t)= i\omega D W(t)$ and so
\beq
	W(t)= e^{i\omega D t}W(0)\ . \label{eq:W-solution}
\eeq
Since $D$ is diagonal it follows that the components $W^J(t)$ of the new vector $W(t)$ evolve in time by 
simply being multiplied by a phase $e^{i d_J \omega t}$, where $d_J$ are the elements on the diagonal of $D$ (i.e., the 
eigenvalues of $M$ which are $0,\pm 2$, and $\pm 4$). The components $W^J(t)$ are all linear combinations of the original 
collective variables $\hat{g}^{\al_1\al_2\al_3\al_4}(t)$, and we can think of them as a new set of collective variables with 
especially simple time-dependence. The presence of the frequency $4\omega$ shows that the quench has indeed excited 
higher-spin collective variables with angular momentum $\pm 4 \hbar$.

The reader may wonder about a certain difference between our present study of higher-spin excitations in the CSMM and the
previous numerical study of higher-spin excitations in Ref.~\onlinecite{LGP}. In the CSMM we find that the matrix $M$ 
discussed above has eigenvalues $0,\pm 2$, and $\pm 4$, indicating that excitations with angular momentum $0,\pm 2\hbar$, and $\pm 4\hbar$ 
are excited by the quench. On the other hand, in Ref.~\onlinecite{LGP} the authors investigated a quench involving the anisotropic
Haldane pseudopotential $\hat{V}_{0,4}$ and found that modes with angular momentum $\pm2\hbar$ were \emph{not} excited,
but higher-spin modes were. The difference
between these two studies is the following. In Ref.~\onlinecite{LGP} the pseudopotential $\hat{V}_{0,4}$ has an octopolar structure
in momentum space (see their Fig.~2b), indicating that $\hat{V}_{0,4}$ excites a pure angular momentum $\pm 4\hbar$ mode. Therefore
in a quench driven by the introduction of $\hat{V}_{0,4}$ one expects to only see modes with angular momentum that is a multiple of
$\pm 4\hbar$. On the other hand, the anisotropy that we introduce in the geometric quench in the CSMM, which is parametrized by 
$g_{ab}$, has a dipolar structure, and it excites angular momentum $\pm 2$ modes. As we can see from 
Eq.~\eqref{eq:post-quench-state}, in the CSMM the post-quench state $|\psi(t)\ran$ is a superposition of states with all possible numbers of spin-2 
quanta excited, and so this state has non-zero overlap with states of any even angular momentum. This is why the quench that we considered
in the CSMM is capable of exciting modes with angular momentum $\pm 2\hbar$, $\pm 4\hbar$, etc.

One final comment is in order regarding the dynamics of these higher-spin observables. The initial values
$W^J(0)$ of the components of $W(t)$ are determined by the initial values $V^J(0)$, which are in turn determined by 
$\hat{g}^{\al_1\al_2\al_3\al_4}(0)$. It follows that if $W^J(0)=0$ for a particular $J$, then  Eq.~\eqref{eq:W-solution} 
implies that $W^J(t)=0$ for all time. 

Let us assume that we have ordered the eigenvectors of $M$ in $S$ in such a way that $d_1= 4$ and $d_2=-4$. Then
the components $W^1(t)$ and $W^2(t)$ evolve in time with the phase factors $e^{i4\omega t}$ and $e^{-i4\omega t}$,
respectively. We would like to check that $W^1(0)$ and $W^2(0)$ are not both zero. If they \emph{were} both
zero, then we would have $W^1(t)=W^2(t)=0$ for all time and we could not legitimately claim that the geometric quench had 
excited the collective variables with spin $4$. 

We now perform a simple check which gives evidence that $W^1(0)$ and $W^2(0)$ are not zero. Specifically, we will check this 
for the case $N=1$ (i.e., the matrix model with one-component matrices). In this case we just have $Z^{-}=z_0$, 
$Z^{+}=z_0^{\dg}$, and the normalized ground state $|\psi_0\ran$ takes the form
\beq
	|\psi_0\ran=\frac{1}{\sqrt{(m-1)!}} (b^{\dg}_1)^{m-1}|0\ran\ ,
\eeq
where $b^{\dg}_1= \frac{1}{\sqrt{h}}\ov{\vphi}_1$ is proportional to the single component of the row vector $\ov{\vphi}^T$, and 
$|0\ran$ is again the Fock vacuum satisfying $z_0|0\ran= b_1|0\ran=0$. We find that in this initial state the only
non-zero components of $V(0)$ are $V^{11}(0)= 1$ and $V^{13}(0)=2$. We have checked numerically for several
values of the anisotropy parameter $A$ that $W^1(0)$ and $W^2(0)$ are not zero in this case. 
Since we do not expect any sudden changes in the properties of the CSMM when we increase
$N$ to values $N>1$, we believe that this check is good evidence that $W^1(0)$ and $W^2(0)$ are not zero
for the CSMM with $N>1$, and so we expect that the geometric quench really does excite these spin 4 observables in the 
CSMM.

%%%%%%%%%%%%%%%%%%%%%%%%%%%%%%%%

\section{Conclusion}
\label{sec:conclusion}

We have investigated the geometric quench protocol for FQH states proposed in Ref.~\onlinecite{LGP} in the context
of exactly solvable matrix models of the Laughlin and Blok-Wen FQH states~\cite{P1,tong2016,dorey2016matrix}. 
We were able to leverage the algebraic
properties of these models to solve the quench exactly, and we then compared the exact solution to previous
results obtained using the bimetric theory of FQH states. Our exact result for the post-quench dynamics of the spin-2 
collective variable $\hat{g}_{ab}(t)$ in the matrix models agrees with the results of bimetric theory in the case of small 
anisotropy, and we 
also showed how the bimetric theory Lagrangian could be altered so that the matrix models and bimetric theory results match
\emph{exactly} for any anisotropy. Beyond the comparison with bimetric theory, we also presented an exact calculation for
the quantum fidelity $|\lan\psi_0|\psi(t)\ran|^2$ after the geometric quench in the matrix models, 
and the expression that we derived seems to be in good agreement with preliminary results of numerical simulations of the 
geometric quench~\cite{ZZ-email}. We also initiated an investigation of the dynamics of higher-spin observables in the 
matrix models, and we showed that the geometric quench leads to a nontrivial dynamics for those observables. Our results here also
give further confirmation for the general picture put forward by two of us in Ref.~\onlinecite{lapa2018}, which is that
quantum Hall matrix models are capable of describing geometric properties of FQH states which are of current interest. 

The major open problem that was partially addressed in the present paper is the dynamics of the higher-spin collective modes. It is clear both from numerical work of Ref.~\onlinecite{LGP} and the present considerations that there are well-defined collective modes of higher angular momentum in FQH states. However, the theoretical description of these modes is plagued by the technical difficulties which we have 
reviewed in Appendix~\ref{app:W}. Presently it is not clear what is the fundamental origin of these difficulties. Development of a unified approach to the higher-spin modes in the language of quantum Hall matrix models, effective field theory, and trial quantum Hall states is an important open problem. 

It is also important to generalize the matrix model description of FQH states to the paired states of Moore-Read~\cite{moore-read} and 
Read-Rezayi~\cite{read-rezayi1999}. These are major candidates for the real-world realization of non-Abelian topological order. Consequently, developing solvable microscopic models that capture both topological and geometric features of these states is an important unsolved problem. 

Finally, electrons in a magnetic field support a variety of spatially-ordered phases known as quantum Hall liquid 
crystals~\cite{fradkin1999liquid}. It would be interesting to implement these phases within the matrix model framework or, more generally, 
in the framework of noncommutative fluids. This possibility is particularly intriguing since both bimetric theory and general noncommutative scalar field theories \cite{gubser2001phase} support spatially-ordered phases.

\acknowledgements

M.F.L. and A.G. acknowledge the support of the Kadanoff Center for Theoretical Physics at the University of Chicago.
A.G. was also supported by the University of Chicago Materials Research Science and Engineering Center, 
which is funded by the National Science Foundation under award number DMR-1420709 and by the Quantum Materials program at LBNL, 
funded by the U.S. Department of Energy under Contract No. DE-AC02-05CH11231.
M.F.L and T.L.H acknowledge support from the U.S. National
Science Foundation under grant DMR 1351895-CAR, as well as the support of the Institute
for Condensed Matter Theory at the University of Illinois at Urbana-Champaign.

\appendix

\section{Some useful formulas for $SU(1,1)$}
\label{app:SU11}
%edited

In this appendix we present several ``rearrangement" identities for exponentials of the $su(1,1)$ generators 
$K_{\pm}$ and $K_0$. We use these identities in Sec.~\ref{sec:quench-solution} of the main text to solve the geometric 
quench in the CSMM. These identities are essentially the same as those used to manipulate squeezed coherent states
of harmonic oscillators (see, for example, Ref.~\onlinecite{Perelomov}).

The first rearrangement identity is
\begin{eqnarray}
	e^{-i\omega t(\sinh(A) K_{+}+2\cosh(A) K_0+\sinh(A) K_{-})} &= \nnb \\
 e^{-\beta(t) K_{+}}e^{\ln(\delta(t)) K_0} &e^{ -\beta(t)K_{-}} \label{eq:id1}
\end{eqnarray}
with
\beqa
	\beta(t) &=& \frac{\sinh(A)}{\cosh(A)-i\cot(\omega t)} \label{eq:beta} \\
	\delta(t) &=& \frac{1}{[\cos(\omega t) + i \cosh(A)\sin(\omega t)]^2}\ . \label{eq:delta}
\eeqa
In addition, in this case the functions $\beta(t)$ and $\delta(t)$ obey the relation (an overline denotes complex conjugation)
\beq
	\frac{\sqrt{\delta(t)\ov{\delta(t)}}}{1-|\beta(t)|^2}= 1\ . \label{eq:id1-extra}
\eeq

The second rearrangement identity is
\beq
	e^{a K_{-}}e^{b K_0} e^{c K_{+}}= e^{a' K_{+}}e^{\ln(b') K_0} e^{c' K_{-}} \label{eq:id2}
\eeq
where $a',b',c'$ are functions of $a,b,c$ and are given explicitly by
\beqa
	a'(a,b,c) &=& \frac{c e^b}{1-ac e^b} \\ 
	b'(a,b,c) &=& \frac{e^b}{(1-ace^b)^2} \label{eq:b-prime} \\
	c'(a,b,c) &=&  \frac{ae^b}{1-ace^b}\ .
\eeqa

The trick to proving these identities is to explicitly check them in a specific representation of $SU(1,1)$ which is easy 
to work with. They are then guaranteed to hold in any other representation (since the operators obey the same algebra in any
representation). The specific representation we use to check these
is the (non-unitary) $2\times 2$ representation in which $K_0= \frac{1}{2}\sigma^z$ and
\beqa
	K_{+} &=& \begin{pmatrix} 
	0 & 1 \\ 
	0 & 0 
\end{pmatrix} \\
K_{-} &=& \begin{pmatrix} 
	0 & 0 \\ 
	-1 & 0 
\end{pmatrix}\ .
\eeqa

\section{Details of the calculations for Sec.~\ref{sec:higher-spin}}
\label{app:big-formulas}
%edited

Here we give the explicit formulas for the vector $V(t)$ and matrix $M$ used in Sec.~\ref{sec:higher-spin}. The
components of $V(t)$ are defined as
\begin{widetext}
\begin{subequations}
\label{eq:V}
\beqa
	V^1(t) &=& \hat{g}^{++++}(t) \\
	V^2(t) &=&  \hat{g}^{+++-}(t) \\
	V^3(t) &=&  \hat{g}^{++-+}(t) \\
	V^4(t) &=&  \hat{g}^{++--}(t)  \\
	V^5(t) &=&  \hat{g}^{+-++}(t) \\
	V^6(t) &=&  \hat{g}^{+-+-}(t)  \\
	V^7(t) &=&  \hat{g}^{+--+}(t) \\
	V^8(t) &=& \hat{g}^{+---}(t) \\
	V^9(t) &=& \hat{g}^{-+++}(t) \\
	V^{10}(t) &=& \hat{g}^{-++-}(t) \\
	V^{11}(t) &=& \hat{g}^{-+-+}(t) \\
	V^{12}(t) &=& \hat{g}^{-+--}(t) \\
	V^{13}(t) &=& \hat{g}^{--++}(t) \\
	V^{14}(t) &=& \hat{g}^{--+-}(t) \\
	V^{15}(t) &=& \hat{g}^{---+}(t) \\
	V^{16}(t) &=& \hat{g}^{----}(t)\ .
\eeqa
\end{subequations}
The matrix $M$ has the form 
{\beq
M=\left(
\begin{array}{cccccccccccccccc}
 4 c_{A} & s_{A} & s_{A} & 0 & s_{A} & 0 & 0 & 0 & s_{A} & 0 & 0 & 0 & 0 & 0 & 0 & 0 \\
 -s_{A} & 2 c_{A} & 0 & s_{A} & 0 & s_{A} & 0 & 0 & 0 & s_{A} & 0 & 0 & 0 & 0 & 0 & 0 \\
 -s_{A} & 0 & 2 c_{A} & s_{A} & 0 & 0 & s_{A} & 0 & 0 & 0 & s_{A} & 0 & 0 & 0 & 0 & 0 \\
 0 & -s_{A} & -s_{A} & 0 & 0 & 0 & 0 & s_{A} & 0 & 0 & 0 & s_{A} & 0 & 0 & 0 & 0 \\
 -s_{A} & 0 & 0 & 0 & 2 c_{A} & s_{A} & s_{A} & 0 & 0 & 0 & 0 & 0 & s_{A} & 0 & 0 & 0 \\
 0 & -s_{A} & 0 & 0 & -s_{A} & 0 & 0 & s_{A} & 0 & 0 & 0 & 0 & 0 & s_{A} & 0 & 0 \\
 0 & 0 & -s_{A} & 0 & -s_{A} & 0 & 0 & s_{A} & 0 & 0 & 0 & 0 & 0 & 0 & s_{A} & 0 \\
 0 & 0 & 0 & -s_{A} & 0 & -s_{A} & -s_{A} & -2 c_{A} & 0 & 0 & 0 & 0 & 0 & 0 & 0 & s_{A} \\
 -s_{A} & 0 & 0 & 0 & 0 & 0 & 0 & 0 & 2 c_{A} & s_{A} & s_{A} & 0 & s_{A} & 0 & 0 & 0 \\
 0 & -s_{A} & 0 & 0 & 0 & 0 & 0 & 0 & -s_{A} & 0 & 0 & s_{A} & 0 & s_{A} & 0 & 0 \\
 0 & 0 & -s_{A} & 0 & 0 & 0 & 0 & 0 & -s_{A} & 0 & 0 & s_{A} & 0 & 0 & s_{A} & 0 \\
 0 & 0 & 0 & -s_{A} & 0 & 0 & 0 & 0 & 0 & -s_{A} & -s_{A} & -2 c_{A} & 0 & 0 & 0 & s_{A} \\
 0 & 0 & 0 & 0 & -s_{A} & 0 & 0 & 0 & -s_{A} & 0 & 0 & 0 & 0 & s_{A} & s_{A} & 0 \\
 0 & 0 & 0 & 0 & 0 & -s_{A} & 0 & 0 & 0 & -s_{A} & 0 & 0 & -s_{A} & -2 c_{A} & 0 & s_{A} \\
 0 & 0 & 0 & 0 & 0 & 0 & -s_{A} & 0 & 0 & 0 & -s_{A} & 0 & -s_{A} & 0 & -2 c_{A} & s_{A} \\
 0 & 0 & 0 & 0 & 0 & 0 & 0 & -s_{A} & 0 & 0 & 0 & -s_{A} & 0 & -s_{A} & -s_{A} & -4 c_{A} \\
\end{array}
\right)\ , \label{eq:M}
\eeq}
where to save space we used a shorthand notation $s_A := \sinh(A)$ and $c_A := \cosh(A)$.
\end{widetext}

%%%%%%%%%
\section{On the commutation relations for the higher-spin operators in the CSMM}
\label{app:W}
%%%%%%%%%

In this appendix we comment on how the operators
$K^{\al_1\al_2\al_3\al_4}$ that we introduced in Sec.~\ref{sec:higher-spin}
are related to previous work on the $W_{\infty}$ algebra in the CSMM \cite{cappelli2005}. The 
authors of Ref.~\onlinecite{cappelli2005} considered higher-spin operators $\mathcal{O}_{n,m}$ in the matrix model
of the form
\beq
	\mathcal{O}_{n,m} = \text{Tr}\{ (Z^{+})^{n+1} (Z^{-})^{m+1}\} \ ,
\eeq
for $m,n\geq -1$. For reasons that we explain below, they found it necessary to also include operators $\mathcal{P}_{n,m}$ 
which depend on the vector $\vphi$ and which are defined as
\beq
	\mathcal{P}_{n,m} = \ov{\vphi}^T (Z^{+})^{n+1} (Z^{-})^{m+1} \vphi\ ,
\eeq
where again we always have $n,m\geq -1$. For any $n<m$ one can show that $\mathcal{O}_{n,m}$ and 
$\mathcal{P}_{n.m}$ annihilate the ground state $|\psi_0\ran$ of the CSMM (the proof is identical to our proof in 
Sec.~\ref{sec:CSMM} that $K_{-}|\psi_0\ran=0$). This
fact, which expresses the incompressibility of the CSMM ground state, is one piece of evidence that these
operators generate the $W_{\infty}$ algebra in the CSMM. However, the algebra obeyed by these operators is not 
exactly the $W_{\infty}$ algebra, and the authors of Ref.~\onlinecite{cappelli2005} were unable to identify a set of 
operators in the CSMM which obey the $W_{\infty}$ algebra exactly. 

To understand what goes wrong in the algebra of these operators it is useful to study a specific example. We consider
the commutator 
\beq
	[\mathcal{O}_{0,2},\mathcal{O}_{1,1}]= [K^{+---},K^{++--} ]\ .
\eeq
When evaluating this commutator one finds many different terms. In some of these terms the quantum operators
and the matrix indices are in the correct order so that the term can be expressed in terms of the original operators 
$\mathcal{O}_{n,m}$. For example we find a term proportional to $\mathcal{O}_{1,3}= K^{++----}$.
In other terms the matrix indices are in the correct order so that the term can be expressed as a trace, but the operators
$Z^{+}$ and $Z^{-}$ (whose matrix elements do not commute as quantum operators) are in the wrong order for the operator 
to be identified with one of the $\mathcal{O}_{n,m}$. For example we find a term proportional to 
$K^{+--+--}$. Finally, we find a third kind of term in which both the matrix ordering and the quantum ordering
prevent one from writing the term in terms of any of the operators we previously defined. For example we find a term
of the form
\beq
	{(Z^{+})^i}_j{\{(Z^{-})^2\}^k}_i {\{(Z^{+}(Z^{-})^2\}^{j}}_k\ ,
\eeq
in which the ordering of the quantum operators clashes with the matrix ordering so that the term cannot be 
identified with any of the operators $K^{\al_1\cdots\al_6}$ or $\mathcal{O}_{n,m}$. 
In Ref.~\onlinecite{cappelli2005} the authors proposed that within the physical Hilbert space of the
CSMM the constraint \eqref{eq:constraint} could be used to simplify complicated terms like this one which arise
in commutators of the $\mathcal{O}_{n,m}$. After using the CSMM constraint one finds 
that the commutator of two $\mathcal{O}_{n,m}$ operators now contains terms involving the 
$\mathcal{P}_{n,m}$ operators,
and this is why the authors of Ref.~\onlinecite{cappelli2005} introduced the $\mathcal{P}_{n,m}$ operators in the
first place. It was conjectured in Ref.~\onlinecite{cappelli2005} that a proper linear combination of $\mathcal{O}_{n,m}$ 
and $\mathcal{P}_{n,m}$ should satisfy the $W_\infty$ algebra exactly.

%\bibliography{quench-refs}

\begin{thebibliography}{74}%
\makeatletter
\providecommand \@ifxundefined [1]{%
 \@ifx{#1\undefined}
}%
\providecommand \@ifnum [1]{%
 \ifnum #1\expandafter \@firstoftwo
 \else \expandafter \@secondoftwo
 \fi
}%
\providecommand \@ifx [1]{%
 \ifx #1\expandafter \@firstoftwo
 \else \expandafter \@secondoftwo
 \fi
}%
\providecommand \natexlab [1]{#1}%
\providecommand \enquote  [1]{``#1''}%
\providecommand \bibnamefont  [1]{#1}%
\providecommand \bibfnamefont [1]{#1}%
\providecommand \citenamefont [1]{#1}%
\providecommand \href@noop [0]{\@secondoftwo}%
\providecommand \href [0]{\begingroup \@sanitize@url \@href}%
\providecommand \@href[1]{\@@startlink{#1}\@@href}%
\providecommand \@@href[1]{\endgroup#1\@@endlink}%
\providecommand \@sanitize@url [0]{\catcode `\\12\catcode `\$12\catcode
  `\&12\catcode `\#12\catcode `\^12\catcode `\_12\catcode `\%12\relax}%
\providecommand \@@startlink[1]{}%
\providecommand \@@endlink[0]{}%
\providecommand \url  [0]{\begingroup\@sanitize@url \@url }%
\providecommand \@url [1]{\endgroup\@href {#1}{\urlprefix }}%
\providecommand \urlprefix  [0]{URL }%
\providecommand \Eprint [0]{\href }%
\providecommand \doibase [0]{http://dx.doi.org/}%
\providecommand \selectlanguage [0]{\@gobble}%
\providecommand \bibinfo  [0]{\@secondoftwo}%
\providecommand \bibfield  [0]{\@secondoftwo}%
\providecommand \translation [1]{[#1]}%
\providecommand \BibitemOpen [0]{}%
\providecommand \bibitemStop [0]{}%
\providecommand \bibitemNoStop [0]{.\EOS\space}%
\providecommand \EOS [0]{\spacefactor3000\relax}%
\providecommand \BibitemShut  [1]{\csname bibitem#1\endcsname}%
\let\auto@bib@innerbib\@empty
%</preamble>
\bibitem [{\citenamefont {Wen}(2004)}]{wen-book}%
  \BibitemOpen
  \bibfield  {author} {\bibinfo {author} {\bibfnamefont {X.-G.}\ \bibnamefont
  {Wen}},\ }\href@noop {} {\emph {\bibinfo {title} {Quantum field theory of
  many-body systems: from the origin of sound to an origin of light and
  electrons}}}\ (\bibinfo  {publisher} {Oxford University Press on Demand},\
  \bibinfo {year} {2004})\BibitemShut {NoStop}%
\bibitem [{\citenamefont {Girvin}\ \emph {et~al.}(1986)\citenamefont {Girvin},
  \citenamefont {MacDonald},\ and\ \citenamefont {Platzman}}]{GMP}%
  \BibitemOpen
  \bibfield  {author} {\bibinfo {author} {\bibfnamefont {S.~M.}\ \bibnamefont
  {Girvin}}, \bibinfo {author} {\bibfnamefont {A.~H.}\ \bibnamefont
  {MacDonald}}, \ and\ \bibinfo {author} {\bibfnamefont {P.~M.}\ \bibnamefont
  {Platzman}},\ }\href@noop {} {\bibfield  {journal} {\bibinfo  {journal}
  {Phys. Rev. B}\ }\textbf {\bibinfo {volume} {33}},\ \bibinfo {pages} {2481}
  (\bibinfo {year} {1986})}\BibitemShut {NoStop}%
\bibitem [{\citenamefont {Gromov}\ \emph {et~al.}(2017)\citenamefont {Gromov},
  \citenamefont {Geraedts},\ and\ \citenamefont {Bradlyn}}]{GGB}%
  \BibitemOpen
  \bibfield  {author} {\bibinfo {author} {\bibfnamefont {A.}~\bibnamefont
  {Gromov}}, \bibinfo {author} {\bibfnamefont {S.~D.}\ \bibnamefont
  {Geraedts}}, \ and\ \bibinfo {author} {\bibfnamefont {B.}~\bibnamefont
  {Bradlyn}},\ }\href@noop {} {\bibfield  {journal} {\bibinfo  {journal} {Phys.
  Rev. Lett.}\ }\textbf {\bibinfo {volume} {119}},\ \bibinfo {pages} {146602}
  (\bibinfo {year} {2017})}\BibitemShut {NoStop}%
\bibitem [{\citenamefont {Gromov}\ and\ \citenamefont
  {Son}(2017)}]{gromov-son}%
  \BibitemOpen
  \bibfield  {author} {\bibinfo {author} {\bibfnamefont {A.}~\bibnamefont
  {Gromov}}\ and\ \bibinfo {author} {\bibfnamefont {D.~T.}\ \bibnamefont
  {Son}},\ }\href@noop {} {\bibfield  {journal} {\bibinfo  {journal} {Phys.
  Rev. X}\ }\textbf {\bibinfo {volume} {7}},\ \bibinfo {pages} {041032}
  (\bibinfo {year} {2017})}\BibitemShut {NoStop}%
\bibitem [{\citenamefont {Abanov}\ and\ \citenamefont {Gromov}(2014)}]{AG}%
  \BibitemOpen
  \bibfield  {author} {\bibinfo {author} {\bibfnamefont {A.~G.}\ \bibnamefont
  {Abanov}}\ and\ \bibinfo {author} {\bibfnamefont {A.}~\bibnamefont
  {Gromov}},\ }\href@noop {} {\bibfield  {journal} {\bibinfo  {journal} {Phys.
  Rev. B}\ }\textbf {\bibinfo {volume} {90}},\ \bibinfo {pages} {014435}
  (\bibinfo {year} {2014})}\BibitemShut {NoStop}%
\bibitem [{\citenamefont {Cho}\ \emph {et~al.}(2014)\citenamefont {Cho},
  \citenamefont {You},\ and\ \citenamefont {Fradkin}}]{cho2014}%
  \BibitemOpen
  \bibfield  {author} {\bibinfo {author} {\bibfnamefont {G.~Y.}\ \bibnamefont
  {Cho}}, \bibinfo {author} {\bibfnamefont {Y.}~\bibnamefont {You}}, \ and\
  \bibinfo {author} {\bibfnamefont {E.}~\bibnamefont {Fradkin}},\ }\href@noop
  {} {\bibfield  {journal} {\bibinfo  {journal} {Phys. Rev. B}\ }\textbf
  {\bibinfo {volume} {90}},\ \bibinfo {pages} {115139} (\bibinfo {year}
  {2014})}\BibitemShut {NoStop}%
\bibitem [{\citenamefont {Ferrari}\ and\ \citenamefont
  {Klevtsov}(2014)}]{ferrari-klevtsov}%
  \BibitemOpen
  \bibfield  {author} {\bibinfo {author} {\bibfnamefont {F.}~\bibnamefont
  {Ferrari}}\ and\ \bibinfo {author} {\bibfnamefont {S.}~\bibnamefont
  {Klevtsov}},\ }\href@noop {} {\bibfield  {journal} {\bibinfo  {journal} {J.
  High Energy Phys.}\ }\textbf {\bibinfo {volume} {2014}},\ \bibinfo {pages}
  {86} (\bibinfo {year} {2014})}\BibitemShut {NoStop}%
\bibitem [{\citenamefont {Bradlyn}\ and\ \citenamefont
  {Read}(2015{\natexlab{a}})}]{BR1}%
  \BibitemOpen
  \bibfield  {author} {\bibinfo {author} {\bibfnamefont {B.}~\bibnamefont
  {Bradlyn}}\ and\ \bibinfo {author} {\bibfnamefont {N.}~\bibnamefont {Read}},\
  }\href@noop {} {\bibfield  {journal} {\bibinfo  {journal} {Phys. Rev. B}\
  }\textbf {\bibinfo {volume} {91}},\ \bibinfo {pages} {165306} (\bibinfo
  {year} {2015}{\natexlab{a}})}\BibitemShut {NoStop}%
\bibitem [{\citenamefont {Bradlyn}\ and\ \citenamefont
  {Read}(2015{\natexlab{b}})}]{BR2}%
  \BibitemOpen
  \bibfield  {author} {\bibinfo {author} {\bibfnamefont {B.}~\bibnamefont
  {Bradlyn}}\ and\ \bibinfo {author} {\bibfnamefont {N.}~\bibnamefont {Read}},\
  }\href@noop {} {\bibfield  {journal} {\bibinfo  {journal} {Phys. Rev. B}\
  }\textbf {\bibinfo {volume} {91}},\ \bibinfo {pages} {125303} (\bibinfo
  {year} {2015}{\natexlab{b}})}\BibitemShut {NoStop}%
\bibitem [{\citenamefont {Can}\ \emph {et~al.}(2015)\citenamefont {Can},
  \citenamefont {Laskin},\ and\ \citenamefont {Wiegmann}}]{CLW}%
  \BibitemOpen
  \bibfield  {author} {\bibinfo {author} {\bibfnamefont {T.}~\bibnamefont
  {Can}}, \bibinfo {author} {\bibfnamefont {M.}~\bibnamefont {Laskin}}, \ and\
  \bibinfo {author} {\bibfnamefont {P.~B.}\ \bibnamefont {Wiegmann}},\
  }\href@noop {} {\bibfield  {journal} {\bibinfo  {journal} {Ann. Phys.}\
  }\textbf {\bibinfo {volume} {362}},\ \bibinfo {pages} {752} (\bibinfo {year}
  {2015})}\BibitemShut {NoStop}%
\bibitem [{\citenamefont {Gromov}\ \emph {et~al.}(2015)\citenamefont {Gromov},
  \citenamefont {Cho}, \citenamefont {You}, \citenamefont {Abanov},\ and\
  \citenamefont {Fradkin}}]{framing}%
  \BibitemOpen
  \bibfield  {author} {\bibinfo {author} {\bibfnamefont {A.}~\bibnamefont
  {Gromov}}, \bibinfo {author} {\bibfnamefont {G.~Y.}\ \bibnamefont {Cho}},
  \bibinfo {author} {\bibfnamefont {Y.}~\bibnamefont {You}}, \bibinfo {author}
  {\bibfnamefont {A.~G.}\ \bibnamefont {Abanov}}, \ and\ \bibinfo {author}
  {\bibfnamefont {E.}~\bibnamefont {Fradkin}},\ }\href@noop {} {\bibfield
  {journal} {\bibinfo  {journal} {Phys. Rev. Lett.}\ }\textbf {\bibinfo
  {volume} {114}},\ \bibinfo {pages} {016805} (\bibinfo {year}
  {2015})}\BibitemShut {NoStop}%
\bibitem [{\citenamefont {Gromov}\ and\ \citenamefont
  {Abanov}(2014)}]{gromov2014density}%
  \BibitemOpen
  \bibfield  {author} {\bibinfo {author} {\bibfnamefont {A.}~\bibnamefont
  {Gromov}}\ and\ \bibinfo {author} {\bibfnamefont {A.~G.}\ \bibnamefont
  {Abanov}},\ }\href@noop {} {\bibfield  {journal} {\bibinfo  {journal} {Phys.
  Rev. Lett.}\ }\textbf {\bibinfo {volume} {113}},\ \bibinfo {pages} {266802}
  (\bibinfo {year} {2014})}\BibitemShut {NoStop}%
\bibitem [{\citenamefont {Karabali}\ and\ \citenamefont {Nair}(2016)}]{KN}%
  \BibitemOpen
  \bibfield  {author} {\bibinfo {author} {\bibfnamefont {D.}~\bibnamefont
  {Karabali}}\ and\ \bibinfo {author} {\bibfnamefont {V.~P.}\ \bibnamefont
  {Nair}},\ }\href@noop {} {\bibfield  {journal} {\bibinfo  {journal} {Phys.
  Rev. D}\ }\textbf {\bibinfo {volume} {94}},\ \bibinfo {pages} {024022}
  (\bibinfo {year} {2016})}\BibitemShut {NoStop}%
\bibitem [{\citenamefont {Haldane}(2009)}]{haldane2009}%
  \BibitemOpen
  \bibfield  {author} {\bibinfo {author} {\bibfnamefont {F.~D.~M.}\
  \bibnamefont {Haldane}},\ }\href@noop {} {\bibfield  {journal} {\bibinfo
  {journal} {arXiv preprint arXiv:0906.1854}\ } (\bibinfo {year}
  {2009})}\BibitemShut {NoStop}%
\bibitem [{\citenamefont {Haldane}(2011)}]{haldane2011}%
  \BibitemOpen
  \bibfield  {author} {\bibinfo {author} {\bibfnamefont {F.~D.~M.}\
  \bibnamefont {Haldane}},\ }\href@noop {} {\bibfield  {journal} {\bibinfo
  {journal} {Phys. Rev. Lett.}\ }\textbf {\bibinfo {volume} {107}},\ \bibinfo
  {pages} {116801} (\bibinfo {year} {2011})}\BibitemShut {NoStop}%
\bibitem [{\citenamefont {Park}\ and\ \citenamefont
  {Haldane}(2014)}]{park-haldane}%
  \BibitemOpen
  \bibfield  {author} {\bibinfo {author} {\bibfnamefont {Y.~J.}\ \bibnamefont
  {Park}}\ and\ \bibinfo {author} {\bibfnamefont {F.~D.~M.}\ \bibnamefont
  {Haldane}},\ }\href@noop {} {\bibfield  {journal} {\bibinfo  {journal} {Phys.
  Rev. B}\ }\textbf {\bibinfo {volume} {90}},\ \bibinfo {pages} {045123}
  (\bibinfo {year} {2014})}\BibitemShut {NoStop}%
\bibitem [{\citenamefont {You}\ \emph {et~al.}(2014)\citenamefont {You},
  \citenamefont {Cho},\ and\ \citenamefont {Fradkin}}]{YCF-nematic}%
  \BibitemOpen
  \bibfield  {author} {\bibinfo {author} {\bibfnamefont {Y.}~\bibnamefont
  {You}}, \bibinfo {author} {\bibfnamefont {G.~Y.}\ \bibnamefont {Cho}}, \ and\
  \bibinfo {author} {\bibfnamefont {E.}~\bibnamefont {Fradkin}},\ }\href@noop
  {} {\bibfield  {journal} {\bibinfo  {journal} {Phys. Rev. X}\ }\textbf
  {\bibinfo {volume} {4}},\ \bibinfo {pages} {041050} (\bibinfo {year}
  {2014})}\BibitemShut {NoStop}%
\bibitem [{\citenamefont {Qiu}\ \emph {et~al.}(2012)\citenamefont {Qiu},
  \citenamefont {Haldane}, \citenamefont {Wan}, \citenamefont {Yang},\ and\
  \citenamefont {Yi}}]{haldane-anisotropic}%
  \BibitemOpen
  \bibfield  {author} {\bibinfo {author} {\bibfnamefont {R.~Z.}\ \bibnamefont
  {Qiu}}, \bibinfo {author} {\bibfnamefont {F.~D.~M.}\ \bibnamefont {Haldane}},
  \bibinfo {author} {\bibfnamefont {X.}~\bibnamefont {Wan}}, \bibinfo {author}
  {\bibfnamefont {K.}~\bibnamefont {Yang}}, \ and\ \bibinfo {author}
  {\bibfnamefont {S.}~\bibnamefont {Yi}},\ }\href@noop {} {\bibfield  {journal}
  {\bibinfo  {journal} {Phys. Rev. B}\ }\textbf {\bibinfo {volume} {85}},\
  \bibinfo {pages} {115308} (\bibinfo {year} {2012})}\BibitemShut {NoStop}%
\bibitem [{\citenamefont {Gromov}\ and\ \citenamefont
  {Abanov}(2015)}]{gromov2015thermal}%
  \BibitemOpen
  \bibfield  {author} {\bibinfo {author} {\bibfnamefont {A.}~\bibnamefont
  {Gromov}}\ and\ \bibinfo {author} {\bibfnamefont {A.~G.}\ \bibnamefont
  {Abanov}},\ }\href@noop {} {\bibfield  {journal} {\bibinfo  {journal} {Phys.
  Rev. Lett.}\ }\textbf {\bibinfo {volume} {114}},\ \bibinfo {pages} {016802}
  (\bibinfo {year} {2015})}\BibitemShut {NoStop}%
\bibitem [{\citenamefont {Gromov}\ \emph {et~al.}(2016)\citenamefont {Gromov},
  \citenamefont {Jensen},\ and\ \citenamefont {Abanov}}]{gromov2016boundary}%
  \BibitemOpen
  \bibfield  {author} {\bibinfo {author} {\bibfnamefont {A.}~\bibnamefont
  {Gromov}}, \bibinfo {author} {\bibfnamefont {K.}~\bibnamefont {Jensen}}, \
  and\ \bibinfo {author} {\bibfnamefont {A.~G.}\ \bibnamefont {Abanov}},\
  }\href@noop {} {\bibfield  {journal} {\bibinfo  {journal} {Phys. Rev. Lett.}\
  }\textbf {\bibinfo {volume} {116}},\ \bibinfo {pages} {126802} (\bibinfo
  {year} {2016})}\BibitemShut {NoStop}%
\bibitem [{\citenamefont {Can}\ \emph {et~al.}(2014)\citenamefont {Can},
  \citenamefont {Laskin},\ and\ \citenamefont {Wiegmann}}]{can2014fractional}%
  \BibitemOpen
  \bibfield  {author} {\bibinfo {author} {\bibfnamefont {T.}~\bibnamefont
  {Can}}, \bibinfo {author} {\bibfnamefont {M.}~\bibnamefont {Laskin}}, \ and\
  \bibinfo {author} {\bibfnamefont {P.}~\bibnamefont {Wiegmann}},\ }\href@noop
  {} {\bibfield  {journal} {\bibinfo  {journal} {Phys. Rev. Lett.}\ }\textbf
  {\bibinfo {volume} {113}},\ \bibinfo {pages} {046803} (\bibinfo {year}
  {2014})}\BibitemShut {NoStop}%
\bibitem [{\citenamefont {Douglas}\ and\ \citenamefont
  {Klevtsov}(2010)}]{douglas2010bergman}%
  \BibitemOpen
  \bibfield  {author} {\bibinfo {author} {\bibfnamefont {M.~R.}\ \bibnamefont
  {Douglas}}\ and\ \bibinfo {author} {\bibfnamefont {S.}~\bibnamefont
  {Klevtsov}},\ }\href@noop {} {\bibfield  {journal} {\bibinfo  {journal}
  {Commun. Math. Phys.}\ }\textbf {\bibinfo {volume} {293}},\ \bibinfo {pages}
  {205} (\bibinfo {year} {2010})}\BibitemShut {NoStop}%
\bibitem [{\citenamefont {Klevtsov}\ and\ \citenamefont
  {Wiegmann}(2015)}]{klevtsov2015geometric}%
  \BibitemOpen
  \bibfield  {author} {\bibinfo {author} {\bibfnamefont {S.}~\bibnamefont
  {Klevtsov}}\ and\ \bibinfo {author} {\bibfnamefont {P.}~\bibnamefont
  {Wiegmann}},\ }\href@noop {} {\bibfield  {journal} {\bibinfo  {journal}
  {Phys. Rev. Lett.}\ }\textbf {\bibinfo {volume} {115}},\ \bibinfo {pages}
  {086801} (\bibinfo {year} {2015})}\BibitemShut {NoStop}%
\bibitem [{\citenamefont {Schine}\ \emph {et~al.}(2016)\citenamefont {Schine},
  \citenamefont {Ryou}, \citenamefont {Gromov}, \citenamefont {Sommer},\ and\
  \citenamefont {Simon}}]{schine2016synthetic}%
  \BibitemOpen
  \bibfield  {author} {\bibinfo {author} {\bibfnamefont {N.}~\bibnamefont
  {Schine}}, \bibinfo {author} {\bibfnamefont {A.}~\bibnamefont {Ryou}},
  \bibinfo {author} {\bibfnamefont {A.}~\bibnamefont {Gromov}}, \bibinfo
  {author} {\bibfnamefont {A.}~\bibnamefont {Sommer}}, \ and\ \bibinfo {author}
  {\bibfnamefont {J.}~\bibnamefont {Simon}},\ }\href@noop {} {\bibfield
  {journal} {\bibinfo  {journal} {Nature}\ }\textbf {\bibinfo {volume} {534}},\
  \bibinfo {pages} {671} (\bibinfo {year} {2016})}\BibitemShut {NoStop}%
\bibitem [{\citenamefont {Schine}\ \emph {et~al.}(2018)\citenamefont {Schine},
  \citenamefont {Chalupnik}, \citenamefont {Can}, \citenamefont {Gromov},\ and\
  \citenamefont {Simon}}]{schine2018measuring}%
  \BibitemOpen
  \bibfield  {author} {\bibinfo {author} {\bibfnamefont {N.}~\bibnamefont
  {Schine}}, \bibinfo {author} {\bibfnamefont {M.}~\bibnamefont {Chalupnik}},
  \bibinfo {author} {\bibfnamefont {T.}~\bibnamefont {Can}}, \bibinfo {author}
  {\bibfnamefont {A.}~\bibnamefont {Gromov}}, \ and\ \bibinfo {author}
  {\bibfnamefont {J.}~\bibnamefont {Simon}},\ }\href@noop {} {\bibfield
  {journal} {\bibinfo  {journal} {arXiv preprint arXiv:1802.04418}\ } (\bibinfo
  {year} {2018})}\BibitemShut {NoStop}%
\bibitem [{\citenamefont {Maciejko}\ \emph {et~al.}(2013)\citenamefont
  {Maciejko}, \citenamefont {Hsu}, \citenamefont {Kivelson}, \citenamefont
  {Park},\ and\ \citenamefont {Sondhi}}]{maciejko2013field}%
  \BibitemOpen
  \bibfield  {author} {\bibinfo {author} {\bibfnamefont {J.}~\bibnamefont
  {Maciejko}}, \bibinfo {author} {\bibfnamefont {B.}~\bibnamefont {Hsu}},
  \bibinfo {author} {\bibfnamefont {S.}~\bibnamefont {Kivelson}}, \bibinfo
  {author} {\bibfnamefont {Y.}~\bibnamefont {Park}}, \ and\ \bibinfo {author}
  {\bibfnamefont {S.}~\bibnamefont {Sondhi}},\ }\href@noop {} {\bibfield
  {journal} {\bibinfo  {journal} {Phys. Rev. B}\ }\textbf {\bibinfo {volume}
  {88}},\ \bibinfo {pages} {125137} (\bibinfo {year} {2013})}\BibitemShut
  {NoStop}%
\bibitem [{\citenamefont {Avron}\ \emph {et~al.}(1995)\citenamefont {Avron},
  \citenamefont {Seiler},\ and\ \citenamefont {Zograf}}]{ASZ}%
  \BibitemOpen
  \bibfield  {author} {\bibinfo {author} {\bibfnamefont {J.~E.}\ \bibnamefont
  {Avron}}, \bibinfo {author} {\bibfnamefont {R.}~\bibnamefont {Seiler}}, \
  and\ \bibinfo {author} {\bibfnamefont {P.~G.}\ \bibnamefont {Zograf}},\
  }\href@noop {} {\bibfield  {journal} {\bibinfo  {journal} {Phys. Rev. Lett.}\
  }\textbf {\bibinfo {volume} {75}},\ \bibinfo {pages} {697} (\bibinfo {year}
  {1995})}\BibitemShut {NoStop}%
\bibitem [{\citenamefont {L{\'e}vay}(1995)}]{levay}%
  \BibitemOpen
  \bibfield  {author} {\bibinfo {author} {\bibfnamefont {P.}~\bibnamefont
  {L{\'e}vay}},\ }\href@noop {} {\bibfield  {journal} {\bibinfo  {journal} {J.
  Math. Phys.}\ }\textbf {\bibinfo {volume} {36}},\ \bibinfo {pages} {2792}
  (\bibinfo {year} {1995})}\BibitemShut {NoStop}%
\bibitem [{\citenamefont {Avron}(1998)}]{avron1998odd}%
  \BibitemOpen
  \bibfield  {author} {\bibinfo {author} {\bibfnamefont {J.~E.}\ \bibnamefont
  {Avron}},\ }\href@noop {} {\bibfield  {journal} {\bibinfo  {journal} {J.
  Stat. Phys.}\ }\textbf {\bibinfo {volume} {92}},\ \bibinfo {pages} {543}
  (\bibinfo {year} {1998})}\BibitemShut {NoStop}%
\bibitem [{\citenamefont {Tokatly}\ and\ \citenamefont {Vignale}(2007)}]{TV1}%
  \BibitemOpen
  \bibfield  {author} {\bibinfo {author} {\bibfnamefont {I.~V.}\ \bibnamefont
  {Tokatly}}\ and\ \bibinfo {author} {\bibfnamefont {G.}~\bibnamefont
  {Vignale}},\ }\href@noop {} {\bibfield  {journal} {\bibinfo  {journal} {Phys.
  Rev. B}\ }\textbf {\bibinfo {volume} {76}},\ \bibinfo {pages} {161305}
  (\bibinfo {year} {2007})}\BibitemShut {NoStop}%
\bibitem [{\citenamefont {Read}(2009)}]{read2009}%
  \BibitemOpen
  \bibfield  {author} {\bibinfo {author} {\bibfnamefont {N.}~\bibnamefont
  {Read}},\ }\href@noop {} {\bibfield  {journal} {\bibinfo  {journal} {Phys.
  Rev. B}\ }\textbf {\bibinfo {volume} {79}},\ \bibinfo {pages} {045308}
  (\bibinfo {year} {2009})}\BibitemShut {NoStop}%
\bibitem [{\citenamefont {Tokatly}\ and\ \citenamefont {Vignale}(2009)}]{TV2}%
  \BibitemOpen
  \bibfield  {author} {\bibinfo {author} {\bibfnamefont {I.~V.}\ \bibnamefont
  {Tokatly}}\ and\ \bibinfo {author} {\bibfnamefont {G.}~\bibnamefont
  {Vignale}},\ }\href@noop {} {\bibfield  {journal} {\bibinfo  {journal} {J.
  Phys. Condens. Matter}\ }\textbf {\bibinfo {volume} {21}},\ \bibinfo {pages}
  {275603} (\bibinfo {year} {2009})}\BibitemShut {NoStop}%
\bibitem [{\citenamefont {Read}\ and\ \citenamefont
  {Rezayi}(2011)}]{read-rezayi}%
  \BibitemOpen
  \bibfield  {author} {\bibinfo {author} {\bibfnamefont {N.}~\bibnamefont
  {Read}}\ and\ \bibinfo {author} {\bibfnamefont {E.~H.}\ \bibnamefont
  {Rezayi}},\ }\href@noop {} {\bibfield  {journal} {\bibinfo  {journal} {Phys.
  Rev. B}\ }\textbf {\bibinfo {volume} {84}},\ \bibinfo {pages} {085316}
  (\bibinfo {year} {2011})}\BibitemShut {NoStop}%
\bibitem [{\citenamefont {Hughes}\ \emph {et~al.}(2011)\citenamefont {Hughes},
  \citenamefont {Leigh},\ and\ \citenamefont {Fradkin}}]{HLF2011}%
  \BibitemOpen
  \bibfield  {author} {\bibinfo {author} {\bibfnamefont {T.~L.}\ \bibnamefont
  {Hughes}}, \bibinfo {author} {\bibfnamefont {R.~G.}\ \bibnamefont {Leigh}}, \
  and\ \bibinfo {author} {\bibfnamefont {E.}~\bibnamefont {Fradkin}},\
  }\href@noop {} {\bibfield  {journal} {\bibinfo  {journal} {Phys. Rev. Lett.}\
  }\textbf {\bibinfo {volume} {107}},\ \bibinfo {pages} {075502} (\bibinfo
  {year} {2011})}\BibitemShut {NoStop}%
\bibitem [{\citenamefont {Hoyos}\ and\ \citenamefont {Son}(2012)}]{hoyos-son}%
  \BibitemOpen
  \bibfield  {author} {\bibinfo {author} {\bibfnamefont {C.}~\bibnamefont
  {Hoyos}}\ and\ \bibinfo {author} {\bibfnamefont {D.~T.}\ \bibnamefont
  {Son}},\ }\href@noop {} {\bibfield  {journal} {\bibinfo  {journal} {Phys.
  Rev. Lett.}\ }\textbf {\bibinfo {volume} {108}},\ \bibinfo {pages} {066805}
  (\bibinfo {year} {2012})}\BibitemShut {NoStop}%
\bibitem [{\citenamefont {Bradlyn}\ \emph {et~al.}(2012)\citenamefont
  {Bradlyn}, \citenamefont {Goldstein},\ and\ \citenamefont
  {Read}}]{bradlyn2012}%
  \BibitemOpen
  \bibfield  {author} {\bibinfo {author} {\bibfnamefont {B.}~\bibnamefont
  {Bradlyn}}, \bibinfo {author} {\bibfnamefont {M.}~\bibnamefont {Goldstein}},
  \ and\ \bibinfo {author} {\bibfnamefont {N.}~\bibnamefont {Read}},\
  }\href@noop {} {\bibfield  {journal} {\bibinfo  {journal} {Phys. Rev. B}\
  }\textbf {\bibinfo {volume} {86}},\ \bibinfo {pages} {245309} (\bibinfo
  {year} {2012})}\BibitemShut {NoStop}%
\bibitem [{\citenamefont {Liu}\ \emph {et~al.}(2018)\citenamefont {Liu},
  \citenamefont {Gromov},\ and\ \citenamefont {Papi\'c}}]{LGP}%
  \BibitemOpen
  \bibfield  {author} {\bibinfo {author} {\bibfnamefont {Z.}~\bibnamefont
  {Liu}}, \bibinfo {author} {\bibfnamefont {A.}~\bibnamefont {Gromov}}, \ and\
  \bibinfo {author} {\bibfnamefont {Z.}~\bibnamefont {Papi\'c}},\ }\href@noop
  {} {\bibfield  {journal} {\bibinfo  {journal} {Phys. Rev. B}\ }\textbf
  {\bibinfo {volume} {98}},\ \bibinfo {pages} {155140} (\bibinfo {year}
  {2018})}\BibitemShut {NoStop}%
\bibitem [{\citenamefont {Yang}\ \emph {et~al.}(2017)\citenamefont {Yang},
  \citenamefont {Hu}, \citenamefont {Lee},\ and\ \citenamefont
  {Papi\ifmmode~\acute{c}\else \'{c}\fi{}}}]{Papic2017}%
  \BibitemOpen
  \bibfield  {author} {\bibinfo {author} {\bibfnamefont {B.}~\bibnamefont
  {Yang}}, \bibinfo {author} {\bibfnamefont {Z.-X.}\ \bibnamefont {Hu}},
  \bibinfo {author} {\bibfnamefont {C.~H.}\ \bibnamefont {Lee}}, \ and\
  \bibinfo {author} {\bibfnamefont {Z.}~\bibnamefont
  {Papi\ifmmode~\acute{c}\else \'{c}\fi{}}},\ }\href@noop {} {\bibfield
  {journal} {\bibinfo  {journal} {Phys. Rev. Lett.}\ }\textbf {\bibinfo
  {volume} {118}},\ \bibinfo {pages} {146403} (\bibinfo {year}
  {2017})}\BibitemShut {NoStop}%
\bibitem [{\citenamefont {Cappelli}\ \emph {et~al.}(1993)\citenamefont
  {Cappelli}, \citenamefont {Trugenberger},\ and\ \citenamefont {Zemba}}]{CTZ}%
  \BibitemOpen
  \bibfield  {author} {\bibinfo {author} {\bibfnamefont {A.}~\bibnamefont
  {Cappelli}}, \bibinfo {author} {\bibfnamefont {C.~A.}\ \bibnamefont
  {Trugenberger}}, \ and\ \bibinfo {author} {\bibfnamefont {G.~R.}\
  \bibnamefont {Zemba}},\ }\href@noop {} {\bibfield  {journal} {\bibinfo
  {journal} {Nucl. Phys. B}\ }\textbf {\bibinfo {volume} {396}},\ \bibinfo
  {pages} {465} (\bibinfo {year} {1993})}\BibitemShut {NoStop}%
\bibitem [{\citenamefont {Karabali}(1994{\natexlab{a}})}]{karabali1994}%
  \BibitemOpen
  \bibfield  {author} {\bibinfo {author} {\bibfnamefont {D.}~\bibnamefont
  {Karabali}},\ }\href@noop {} {\bibfield  {journal} {\bibinfo  {journal}
  {Nucl. Phys. B}\ }\textbf {\bibinfo {volume} {419}},\ \bibinfo {pages} {437}
  (\bibinfo {year} {1994}{\natexlab{a}})}\BibitemShut {NoStop}%
\bibitem [{\citenamefont {Karabali}(1994{\natexlab{b}})}]{karabali1994-2}%
  \BibitemOpen
  \bibfield  {author} {\bibinfo {author} {\bibfnamefont {D.}~\bibnamefont
  {Karabali}},\ }\href@noop {} {\bibfield  {journal} {\bibinfo  {journal}
  {Nucl. Phys. B}\ }\textbf {\bibinfo {volume} {428}},\ \bibinfo {pages} {531}
  (\bibinfo {year} {1994}{\natexlab{b}})}\BibitemShut {NoStop}%
\bibitem [{\citenamefont {Flohr}\ and\ \citenamefont
  {Varnhagen}(1994)}]{flohr1994}%
  \BibitemOpen
  \bibfield  {author} {\bibinfo {author} {\bibfnamefont {M.}~\bibnamefont
  {Flohr}}\ and\ \bibinfo {author} {\bibfnamefont {R.}~\bibnamefont
  {Varnhagen}},\ }\href@noop {} {\bibfield  {journal} {\bibinfo  {journal} {J.
  Phys. A}\ }\textbf {\bibinfo {volume} {27}},\ \bibinfo {pages} {3999}
  (\bibinfo {year} {1994})}\BibitemShut {NoStop}%
\bibitem [{\citenamefont {Cappelli}\ \emph {et~al.}(1994)\citenamefont
  {Cappelli}, \citenamefont {Trugenberger},\ and\ \citenamefont
  {Zemba}}]{CTZ-2}%
  \BibitemOpen
  \bibfield  {author} {\bibinfo {author} {\bibfnamefont {A.}~\bibnamefont
  {Cappelli}}, \bibinfo {author} {\bibfnamefont {C.~A.}\ \bibnamefont
  {Trugenberger}}, \ and\ \bibinfo {author} {\bibfnamefont {G.~R.}\
  \bibnamefont {Zemba}},\ }\href@noop {} {\bibfield  {journal} {\bibinfo
  {journal} {Phys. Rev. Lett.}\ }\textbf {\bibinfo {volume} {72}},\ \bibinfo
  {pages} {1902} (\bibinfo {year} {1994})}\BibitemShut {NoStop}%
\bibitem [{\citenamefont {Polychronakos}(2001{\natexlab{a}})}]{P1}%
  \BibitemOpen
  \bibfield  {author} {\bibinfo {author} {\bibfnamefont {A.~P.}\ \bibnamefont
  {Polychronakos}},\ }\href@noop {} {\bibfield  {journal} {\bibinfo  {journal}
  {J. High Energy Phys.}\ }\textbf {\bibinfo {volume} {2001}},\ \bibinfo
  {pages} {011} (\bibinfo {year} {2001}{\natexlab{a}})}\BibitemShut {NoStop}%
\bibitem [{\citenamefont {Susskind}(2001)}]{susskind}%
  \BibitemOpen
  \bibfield  {author} {\bibinfo {author} {\bibfnamefont {L.}~\bibnamefont
  {Susskind}},\ }\href@noop {} {\bibfield  {journal} {\bibinfo  {journal}
  {arXiv preprint hep-th/0101029}\ } (\bibinfo {year} {2001})}\BibitemShut
  {NoStop}%
\bibitem [{\citenamefont {Polychronakos}(2001{\natexlab{b}})}]{P3}%
  \BibitemOpen
  \bibfield  {author} {\bibinfo {author} {\bibfnamefont {A.~P.}\ \bibnamefont
  {Polychronakos}},\ }\href@noop {} {\bibfield  {journal} {\bibinfo  {journal}
  {J. High Energy Phys.}\ }\textbf {\bibinfo {volume} {2001}},\ \bibinfo
  {pages} {070} (\bibinfo {year} {2001}{\natexlab{b}})}\BibitemShut {NoStop}%
\bibitem [{\citenamefont {Morariu}\ and\ \citenamefont
  {Polychronakos}(2001)}]{MP}%
  \BibitemOpen
  \bibfield  {author} {\bibinfo {author} {\bibfnamefont {B.}~\bibnamefont
  {Morariu}}\ and\ \bibinfo {author} {\bibfnamefont {A.~P.}\ \bibnamefont
  {Polychronakos}},\ }\href@noop {} {\bibfield  {journal} {\bibinfo  {journal}
  {J. High Energy Phys.}\ }\textbf {\bibinfo {volume} {2001}},\ \bibinfo
  {pages} {006} (\bibinfo {year} {2001})}\BibitemShut {NoStop}%
\bibitem [{\citenamefont {Hellerman}\ and\ \citenamefont
  {Van~Raamsdonk}(2001)}]{HVR}%
  \BibitemOpen
  \bibfield  {author} {\bibinfo {author} {\bibfnamefont {S.}~\bibnamefont
  {Hellerman}}\ and\ \bibinfo {author} {\bibfnamefont {M.}~\bibnamefont
  {Van~Raamsdonk}},\ }\href@noop {} {\bibfield  {journal} {\bibinfo  {journal}
  {J. High Energy Phys.}\ }\textbf {\bibinfo {volume} {2001}},\ \bibinfo
  {pages} {039} (\bibinfo {year} {2001})}\BibitemShut {NoStop}%
\bibitem [{\citenamefont {Karabali}\ and\ \citenamefont
  {Sakita}(2001)}]{karabali-sakita1}%
  \BibitemOpen
  \bibfield  {author} {\bibinfo {author} {\bibfnamefont {D.}~\bibnamefont
  {Karabali}}\ and\ \bibinfo {author} {\bibfnamefont {B.}~\bibnamefont
  {Sakita}},\ }\href@noop {} {\bibfield  {journal} {\bibinfo  {journal} {Phys.
  Rev. B}\ }\textbf {\bibinfo {volume} {64}},\ \bibinfo {pages} {245316}
  (\bibinfo {year} {2001})}\BibitemShut {NoStop}%
\bibitem [{\citenamefont {Karabali}\ and\ \citenamefont
  {Sakita}(2002)}]{karabali-sakita2}%
  \BibitemOpen
  \bibfield  {author} {\bibinfo {author} {\bibfnamefont {D.}~\bibnamefont
  {Karabali}}\ and\ \bibinfo {author} {\bibfnamefont {B.}~\bibnamefont
  {Sakita}},\ }\href@noop {} {\bibfield  {journal} {\bibinfo  {journal} {Phys.
  Rev. B}\ }\textbf {\bibinfo {volume} {65}},\ \bibinfo {pages} {075304}
  (\bibinfo {year} {2002})}\BibitemShut {NoStop}%
\bibitem [{\citenamefont {Hansson}\ and\ \citenamefont
  {Karlhede}(2001)}]{hansson2001}%
  \BibitemOpen
  \bibfield  {author} {\bibinfo {author} {\bibfnamefont {T.}~\bibnamefont
  {Hansson}}\ and\ \bibinfo {author} {\bibfnamefont {A.}~\bibnamefont
  {Karlhede}},\ }\href@noop {} {\bibfield  {journal} {\bibinfo  {journal}
  {arXiv preprint cond-mat/0109413}\ } (\bibinfo {year} {2001})}\BibitemShut
  {NoStop}%
\bibitem [{\citenamefont {Fradkin}\ \emph {et~al.}(2002)\citenamefont
  {Fradkin}, \citenamefont {Jejjala},\ and\ \citenamefont
  {Leigh}}]{fradkin-NCCS}%
  \BibitemOpen
  \bibfield  {author} {\bibinfo {author} {\bibfnamefont {E.}~\bibnamefont
  {Fradkin}}, \bibinfo {author} {\bibfnamefont {V.}~\bibnamefont {Jejjala}}, \
  and\ \bibinfo {author} {\bibfnamefont {R.~G.}\ \bibnamefont {Leigh}},\
  }\href@noop {} {\bibfield  {journal} {\bibinfo  {journal} {Nucl. Phys. B}\
  }\textbf {\bibinfo {volume} {642}},\ \bibinfo {pages} {483} (\bibinfo {year}
  {2002})}\BibitemShut {NoStop}%
\bibitem [{\citenamefont {Hansson}\ \emph {et~al.}(2003)\citenamefont
  {Hansson}, \citenamefont {Kailasvuori},\ and\ \citenamefont
  {Karlhede}}]{hansson2003}%
  \BibitemOpen
  \bibfield  {author} {\bibinfo {author} {\bibfnamefont {T.~H.}\ \bibnamefont
  {Hansson}}, \bibinfo {author} {\bibfnamefont {J.}~\bibnamefont
  {Kailasvuori}}, \ and\ \bibinfo {author} {\bibfnamefont {A.}~\bibnamefont
  {Karlhede}},\ }\href@noop {} {\bibfield  {journal} {\bibinfo  {journal}
  {Phys. Rev. B}\ }\textbf {\bibinfo {volume} {68}},\ \bibinfo {pages} {035327}
  (\bibinfo {year} {2003})}\BibitemShut {NoStop}%
\bibitem [{\citenamefont {Cappelli}\ and\ \citenamefont
  {Riccardi}(2005)}]{cappelli2005}%
  \BibitemOpen
  \bibfield  {author} {\bibinfo {author} {\bibfnamefont {A.}~\bibnamefont
  {Cappelli}}\ and\ \bibinfo {author} {\bibfnamefont {M.}~\bibnamefont
  {Riccardi}},\ }\href@noop {} {\bibfield  {journal} {\bibinfo  {journal} {J.
  Stat. Mech. Theor. Exp.}\ }\textbf {\bibinfo {volume} {2005}},\ \bibinfo
  {pages} {P05001} (\bibinfo {year} {2005})}\BibitemShut {NoStop}%
\bibitem [{\citenamefont {Tong}\ and\ \citenamefont
  {Turner}(2015)}]{tong-turner2015}%
  \BibitemOpen
  \bibfield  {author} {\bibinfo {author} {\bibfnamefont {D.}~\bibnamefont
  {Tong}}\ and\ \bibinfo {author} {\bibfnamefont {C.}~\bibnamefont {Turner}},\
  }\href@noop {} {\bibfield  {journal} {\bibinfo  {journal} {Phys. Rev. B}\
  }\textbf {\bibinfo {volume} {92}},\ \bibinfo {pages} {235125} (\bibinfo
  {year} {2015})}\BibitemShut {NoStop}%
\bibitem [{\citenamefont {Blok}\ and\ \citenamefont {Wen}(1992)}]{blok-wen}%
  \BibitemOpen
  \bibfield  {author} {\bibinfo {author} {\bibfnamefont {B.}~\bibnamefont
  {Blok}}\ and\ \bibinfo {author} {\bibfnamefont {X.-G.}\ \bibnamefont {Wen}},\
  }\href@noop {} {\bibfield  {journal} {\bibinfo  {journal} {Nucl. Phys. B}\
  }\textbf {\bibinfo {volume} {374}},\ \bibinfo {pages} {615} (\bibinfo {year}
  {1992})}\BibitemShut {NoStop}%
\bibitem [{\citenamefont {Dorey}\ \emph
  {et~al.}(2016{\natexlab{a}})\citenamefont {Dorey}, \citenamefont {Tong},\
  and\ \citenamefont {Turner}}]{tong2016}%
  \BibitemOpen
  \bibfield  {author} {\bibinfo {author} {\bibfnamefont {N.}~\bibnamefont
  {Dorey}}, \bibinfo {author} {\bibfnamefont {D.}~\bibnamefont {Tong}}, \ and\
  \bibinfo {author} {\bibfnamefont {C.}~\bibnamefont {Turner}},\ }\href@noop {}
  {\bibfield  {journal} {\bibinfo  {journal} {Phys. Rev. B}\ }\textbf {\bibinfo
  {volume} {94}},\ \bibinfo {pages} {085114} (\bibinfo {year}
  {2016}{\natexlab{a}})}\BibitemShut {NoStop}%
\bibitem [{\citenamefont {Dorey}\ \emph
  {et~al.}(2016{\natexlab{b}})\citenamefont {Dorey}, \citenamefont {Tong},\
  and\ \citenamefont {Turner}}]{dorey2016matrix}%
  \BibitemOpen
  \bibfield  {author} {\bibinfo {author} {\bibfnamefont {N.}~\bibnamefont
  {Dorey}}, \bibinfo {author} {\bibfnamefont {D.}~\bibnamefont {Tong}}, \ and\
  \bibinfo {author} {\bibfnamefont {C.}~\bibnamefont {Turner}},\ }\href@noop {}
  {\bibfield  {journal} {\bibinfo  {journal} {J. High Energy Phys.}\ }\textbf
  {\bibinfo {volume} {2016}},\ \bibinfo {pages} {7} (\bibinfo {year}
  {2016}{\natexlab{b}})}\BibitemShut {NoStop}%
\bibitem [{\citenamefont {Lapa}\ and\ \citenamefont {Hughes}(2018)}]{lapa2018}%
  \BibitemOpen
  \bibfield  {author} {\bibinfo {author} {\bibfnamefont {M.~F.}\ \bibnamefont
  {Lapa}}\ and\ \bibinfo {author} {\bibfnamefont {T.~L.}\ \bibnamefont
  {Hughes}},\ }\href@noop {} {\bibfield  {journal} {\bibinfo  {journal} {Phys.
  Rev. B}\ }\textbf {\bibinfo {volume} {97}},\ \bibinfo {pages} {205122}
  (\bibinfo {year} {2018})}\BibitemShut {NoStop}%
\bibitem [{\citenamefont {Lapa}\ \emph {et~al.}(2018)\citenamefont {Lapa},
  \citenamefont {Turner}, \citenamefont {Hughes},\ and\ \citenamefont
  {Tong}}]{LHTT}%
  \BibitemOpen
  \bibfield  {author} {\bibinfo {author} {\bibfnamefont {M.~F.}\ \bibnamefont
  {Lapa}}, \bibinfo {author} {\bibfnamefont {C.}~\bibnamefont {Turner}},
  \bibinfo {author} {\bibfnamefont {T.~L.}\ \bibnamefont {Hughes}}, \ and\
  \bibinfo {author} {\bibfnamefont {D.}~\bibnamefont {Tong}},\ }\href@noop {}
  {\bibfield  {journal} {\bibinfo  {journal} {Phys. Rev. B}\ }\textbf {\bibinfo
  {volume} {98}},\ \bibinfo {pages} {075133} (\bibinfo {year}
  {2018})}\BibitemShut {NoStop}%
\bibitem [{\citenamefont {Rajabpour}\ and\ \citenamefont
  {Sotiriadis}(2014)}]{rajabpour2014}%
  \BibitemOpen
  \bibfield  {author} {\bibinfo {author} {\bibfnamefont {M.}~\bibnamefont
  {Rajabpour}}\ and\ \bibinfo {author} {\bibfnamefont {S.}~\bibnamefont
  {Sotiriadis}},\ }\href@noop {} {\bibfield  {journal} {\bibinfo  {journal}
  {Phys. Rev. A}\ }\textbf {\bibinfo {volume} {89}},\ \bibinfo {pages} {033620}
  (\bibinfo {year} {2014})}\BibitemShut {NoStop}%
\bibitem [{\citenamefont {Franchini}\ \emph {et~al.}(2015)\citenamefont
  {Franchini}, \citenamefont {Gromov}, \citenamefont {Kulkarni},\ and\
  \citenamefont {Trombettoni}}]{franchini2015universal}%
  \BibitemOpen
  \bibfield  {author} {\bibinfo {author} {\bibfnamefont {F.}~\bibnamefont
  {Franchini}}, \bibinfo {author} {\bibfnamefont {A.}~\bibnamefont {Gromov}},
  \bibinfo {author} {\bibfnamefont {M.}~\bibnamefont {Kulkarni}}, \ and\
  \bibinfo {author} {\bibfnamefont {A.}~\bibnamefont {Trombettoni}},\
  }\href@noop {} {\bibfield  {journal} {\bibinfo  {journal} {J. Phys. A}\
  }\textbf {\bibinfo {volume} {48}},\ \bibinfo {pages} {28FT01} (\bibinfo
  {year} {2015})}\BibitemShut {NoStop}%
\bibitem [{\citenamefont {Wiegmann}(2012)}]{wiegmann2012}%
  \BibitemOpen
  \bibfield  {author} {\bibinfo {author} {\bibfnamefont {P.}~\bibnamefont
  {Wiegmann}},\ }\href@noop {} {\bibfield  {journal} {\bibinfo  {journal}
  {PRL}\ }\textbf {\bibinfo {volume} {108}},\ \bibinfo {pages} {206810}
  (\bibinfo {year} {2012})}\BibitemShut {NoStop}%
\bibitem [{\citenamefont {Liu}\ and\ \citenamefont {Papi{\'c}}()}]{ZZ-email}%
  \BibitemOpen
  \bibfield  {author} {\bibinfo {author} {\bibfnamefont {Z.}~\bibnamefont
  {Liu}}\ and\ \bibinfo {author} {\bibfnamefont {Z.}~\bibnamefont
  {Papi{\'c}}},\ }\href@noop {} {}\bibinfo {howpublished} {private
  communication}\BibitemShut {NoStop}%
\bibitem [{\citenamefont {Read}(1998)}]{read1998lowest}%
  \BibitemOpen
  \bibfield  {author} {\bibinfo {author} {\bibfnamefont {N.}~\bibnamefont
  {Read}},\ }\href@noop {} {\bibfield  {journal} {\bibinfo  {journal} {Phys.
  Rev. B}\ }\textbf {\bibinfo {volume} {58}},\ \bibinfo {pages} {16262}
  (\bibinfo {year} {1998})}\BibitemShut {NoStop}%
\bibitem [{\citenamefont {Golkar}\ \emph {et~al.}(2016)\citenamefont {Golkar},
  \citenamefont {Nguyen}, \citenamefont {Roberts},\ and\ \citenamefont
  {Son}}]{golkar2016higher}%
  \BibitemOpen
  \bibfield  {author} {\bibinfo {author} {\bibfnamefont {S.}~\bibnamefont
  {Golkar}}, \bibinfo {author} {\bibfnamefont {D.~X.}\ \bibnamefont {Nguyen}},
  \bibinfo {author} {\bibfnamefont {M.~M.}\ \bibnamefont {Roberts}}, \ and\
  \bibinfo {author} {\bibfnamefont {D.~T.}\ \bibnamefont {Son}},\ }\href@noop
  {} {\bibfield  {journal} {\bibinfo  {journal} {Phys. Rev. Lett.}\ }\textbf
  {\bibinfo {volume} {117}},\ \bibinfo {pages} {216403} (\bibinfo {year}
  {2016})}\BibitemShut {NoStop}%
\bibitem [{\citenamefont {Nguyen}\ \emph {et~al.}(2018)\citenamefont {Nguyen},
  \citenamefont {Gromov},\ and\ \citenamefont {Son}}]{nguyen2018fractional}%
  \BibitemOpen
  \bibfield  {author} {\bibinfo {author} {\bibfnamefont {D.~X.}\ \bibnamefont
  {Nguyen}}, \bibinfo {author} {\bibfnamefont {A.}~\bibnamefont {Gromov}}, \
  and\ \bibinfo {author} {\bibfnamefont {D.~T.}\ \bibnamefont {Son}},\
  }\href@noop {} {\bibfield  {journal} {\bibinfo  {journal} {Phys. Rev. B}\
  }\textbf {\bibinfo {volume} {97}},\ \bibinfo {pages} {195103} (\bibinfo
  {year} {2018})}\BibitemShut {NoStop}%
\bibitem [{\citenamefont {Cappelli}\ and\ \citenamefont
  {Randellini}(2016)}]{cappelli2016multipole}%
  \BibitemOpen
  \bibfield  {author} {\bibinfo {author} {\bibfnamefont {A.}~\bibnamefont
  {Cappelli}}\ and\ \bibinfo {author} {\bibfnamefont {E.}~\bibnamefont
  {Randellini}},\ }\href@noop {} {\bibfield  {journal} {\bibinfo  {journal} {J.
  High Energy Phys.}\ }\textbf {\bibinfo {volume} {2016}},\ \bibinfo {pages}
  {105} (\bibinfo {year} {2016})}\BibitemShut {NoStop}%
\bibitem [{\citenamefont {Inc.}()}]{math}%
  \BibitemOpen
  \bibfield  {author} {\bibinfo {author} {\bibfnamefont {W.~R.}\ \bibnamefont
  {Inc.}},\ }\href@noop {} {\enquote {\bibinfo {title} {Mathematica, {V}ersion
  11.3},}\ }\bibinfo {note} {Champaign, IL, 2018}\BibitemShut {NoStop}%
\bibitem [{\citenamefont {Moore}\ and\ \citenamefont
  {Read}(1991)}]{moore-read}%
  \BibitemOpen
  \bibfield  {author} {\bibinfo {author} {\bibfnamefont {G.}~\bibnamefont
  {Moore}}\ and\ \bibinfo {author} {\bibfnamefont {N.}~\bibnamefont {Read}},\
  }\href@noop {} {\bibfield  {journal} {\bibinfo  {journal} {Nucl. Phys. B}\
  }\textbf {\bibinfo {volume} {360}},\ \bibinfo {pages} {362} (\bibinfo {year}
  {1991})}\BibitemShut {NoStop}%
\bibitem [{\citenamefont {Read}\ and\ \citenamefont
  {Rezayi}(1999)}]{read-rezayi1999}%
  \BibitemOpen
  \bibfield  {author} {\bibinfo {author} {\bibfnamefont {N.}~\bibnamefont
  {Read}}\ and\ \bibinfo {author} {\bibfnamefont {E.}~\bibnamefont {Rezayi}},\
  }\href@noop {} {\bibfield  {journal} {\bibinfo  {journal} {Phys. Rev. B}\
  }\textbf {\bibinfo {volume} {59}},\ \bibinfo {pages} {8084} (\bibinfo {year}
  {1999})}\BibitemShut {NoStop}%
\bibitem [{\citenamefont {Fradkin}\ and\ \citenamefont
  {Kivelson}(1999)}]{fradkin1999liquid}%
  \BibitemOpen
  \bibfield  {author} {\bibinfo {author} {\bibfnamefont {E.}~\bibnamefont
  {Fradkin}}\ and\ \bibinfo {author} {\bibfnamefont {S.~A.}\ \bibnamefont
  {Kivelson}},\ }\href@noop {} {\bibfield  {journal} {\bibinfo  {journal}
  {Phys. Rev. B}\ }\textbf {\bibinfo {volume} {59}},\ \bibinfo {pages} {8065}
  (\bibinfo {year} {1999})}\BibitemShut {NoStop}%
\bibitem [{\citenamefont {Gubser}\ and\ \citenamefont
  {Sondhi}(2001)}]{gubser2001phase}%
  \BibitemOpen
  \bibfield  {author} {\bibinfo {author} {\bibfnamefont {S.~S.}\ \bibnamefont
  {Gubser}}\ and\ \bibinfo {author} {\bibfnamefont {S.~L.}\ \bibnamefont
  {Sondhi}},\ }\href@noop {} {\bibfield  {journal} {\bibinfo  {journal} {Nucl.
  Phys. B}\ }\textbf {\bibinfo {volume} {605}},\ \bibinfo {pages} {395}
  (\bibinfo {year} {2001})}\BibitemShut {NoStop}%
\bibitem [{\citenamefont {Perelomov}(1977)}]{Perelomov}%
  \BibitemOpen
  \bibfield  {author} {\bibinfo {author} {\bibfnamefont {A.~M.}\ \bibnamefont
  {Perelomov}},\ }\href@noop {} {\bibfield  {journal} {\bibinfo  {journal}
  {Physics-Uspekhi}\ }\textbf {\bibinfo {volume} {20}},\ \bibinfo {pages} {703}
  (\bibinfo {year} {1977})}\BibitemShut {NoStop}%
\end{thebibliography}

%merlin.mbs apsrev4-1.bst 2010-07-25 4.21a (PWD, AO, DPC) hacked
%Control: key (0)
%Control: author (8) initials jnrlst
%Control: editor formatted (1) identically to author
%Control: production of article title (-1) disabled
%Control: page (0) single
%Control: year (1) truncated
%Control: production of eprint (0) enabled
%

\end{document}